\documentclass[11pt]{article}
\usepackage{epsf,amsmath,amssymb,graphicx,scalefnt,rotating,enumitem,fancyhdr}
\usepackage[english]{babel}
\usepackage[noadjust]{cite}
\usepackage[final]{showkeys}
\usepackage{slashed}
\usepackage{dsfont}
\usepackage[acronym]{glossaries}
\usepackage{listings}
\let\oldglsentryshort\glsentryshort
\renewcommand{\glsentryshort}[1]{\abbrev{\oldglsentryshort{#1}}}
\usepackage{filemod}
\usepackage[pdftex, colorlinks=true, linkcolor=black, filecolor=black,
  citecolor=black, urlcolor=black, draft=false, bookmarks,
  bookmarksnumbered=true, plainpages=false,
  linktocpage={true}]{hyperref}
\usepackage[affil-it]{authblk}
\usepackage[usenames,dvipsnames]{color}
\usepackage[capitalize]{cleveref}
\newcommand{\lmut}{L_{\mu t}}

\definecolor{codegreen}{rgb}{0,0.6,0}
\definecolor{codegray}{rgb}{0.5,0.5,0.5}
\definecolor{codepurple}{rgb}{0.58,0,0.82}
\definecolor{backcolour}{rgb}{0.95,0.95,0.92}
\lstdefinestyle{mystyle}{
    backgroundcolor=\color{backcolour},
    commentstyle=\color{codegreen},
    keywordstyle=\color{magenta},
    numberstyle=\tiny\color{codegray},
    stringstyle=\color{codepurple},
    basicstyle=\ttfamily\footnotesize,
    breakatwhitespace=false,
    breaklines=true,
    captionpos=b,
    keepspaces=true,
    numbers=left,
    numbersep=5pt,
    showspaces=false,
    showstringspaces=false,
    showtabs=false,
    tabsize=2
}

\newcommand{\nl}{N_\text{L}}
\newcommand{\nf}{N_\text{F}}
\newcommand{\nc}{n_\text{c}}
\newcommand{\na}{n_\text{A}}
\newcommand{\one}{one}
\newcommand{\two}{two}
\newcommand{\three}{three}
\newcommand{\four}{four}
\newcommand{\five}{five}

\newcommand{\wrt}{w.r.t.}

\newcommand{\citere}[1]{Ref.\,\cite{#1}}
\newcommand{\citeres}[1]{Refs.\,\cite{#1}}

\newcommand{\abbrev}[1]{{\scalefont{.9}#1}}
\newcommand{\EulerGamma}{\gamma_\text{E}}

\newcommand{\bare}{0}
\newcommand{\ep}{\epsilon}
\newcommand{\api}{a_{4\pi}}
\newcommand{\dd}{\mathrm{d}}
\newcommand{\deriv}[3]{\frac{\partial\ifthenelse{\equal{#1}{}}{}{^{#1}}%
    #2}{\partial #3\ifthenelse{\equal{#1}{}}{}{^{#1}}}}
\newcommand{\dderiv}[3]{\frac{\dd\ifthenelse{\equal{#1}{}}{}{^{#1}}%
    #2}{\dd #3\ifthenelse{\equal{#1}{}}{}{^{#1}}}}
\newcommand{\order}[1]{{\cal O}(#1)}

\newcommand{\msbar}{\ensuremath{\overline{\mbox{\abbrev{MS}}}}}

\newcommand{\ccf}{C_\text{F}}
\newcommand{\cca}{C_\text{A}}
\newcommand{\ctr}{T_\text{R}}

\newacronym{RG}{RG}{renormalization group}
\newcommand{\rg}{\gls{RG}}
\newacronym{GFF}{GFF}{gradient-flow formalism}
\newcommand{\gff}{\gls{GFF}}
\newacronym{SFTX}{SFTX}{short-flow-time expansion}
\newcommand{\sftx}{\gls{SFTX}}
\newacronym[\glslongpluralkey={degrees of freedom},%
  shortplural=d.o.f.]{dof}{d.o.f.}{degree of freedom}

\newacronym[\glslongpluralkey={renormalization-group equations},%
  shortplural=\abbrev{RGE}s]{RGE}{RGE}{renormalization-group equation}

\newacronym{LSZ}{LSZ}{Lehmann-Symanzik-Zimmermann}

\newacronym[\glslongpluralkey={equation-of-motions},%
  shortplural=\abbrev{EoM}s]{EoM}{EoM}{equation-of-motion}

\newacronym[\glslongpluralkey={Effective Field Theories},%
  shortplural=\abbrev{EFT}s]{EFT}{EFT}{Effective Field Theory}

\newacronym{QED}{QED}{quantum elecrodynamics}

\newacronym{SSB}{SSB}{spontaneous symmetry breaking}

\newacronym[\glslongpluralkey={vacuum expectation values},%
  shortplural=\abbrev{VEV}s]{VEV}{VEV}{vacuum expectation value}
\newcommand{\vev}{\gls{VEV}}
\newcommand{\vevs}{\glspl{VEV}}
\newacronym{QM}{QM}{quantum mechanics}

\newacronym[\glslongpluralkey={quantum field theories},%
  shortplural=\abbrev{QFT}s]{QFT}{QFT}{quantum field theory}

\newacronym{UV}{UV}{ultra-violet}
\newcommand{\uv}{\gls{UV}}
\newcommand{\qcd}{\abbrev{QCD}}
\newacronym{LHC}{LHC}{Large Hadron Collider}

\newacronym{PMNS}{PMNS}{Pontecorvo–Maki–Nakagawa–Sakata}

\newacronym{CKM}{CKM}{Cabbibo-Kobayashi-Maskawa}

\newacronym{IbP}{IbP}{integration-by-parts}

\newacronym{LO}{LO}{leading order}
\newcommand{\lo}{\gls{LO}}
\newacronym{NLO}{NLO}{next-to-leading order}
\newcommand{\nlo}{\gls{NLO}}
\newacronym{NNLO}{NNLO}{next-to-next-to-leading order}
\newcommand{\nnlo}{\gls{NNLO}}
\newacronym{LL}{LL}{leading logarithmic}
\newacronym{NLL}{NLL}{next-to-leading logarithmic}

\newacronym{NNLL}{NNLL}{next-to-next-to-leading logarithmic}

\newacronym[\glslongpluralkey={parton density functions},%
  shortplural=\abbrev{PDF}s]{PDF}{PDF}{parton density function}

\newacronym{SM}{SM}{Standard Model}

\newacronym{SMEFT}{SMEFT}{Standard Model Effective Field Theory}

\newacronym{BSM}{BSM}{beyond-the-\gls{SM}}

\newacronym{MSSM}{MSSM}{Minimal Supersymmetric \gls{SM}}

\newacronym{SUSY}{SUSY}{Supersymmetry}

\newacronym{DREG}{DREG}{Dimensional Regularization}

\newacronym{DRED}{DRED}{Dimensional Reduction}

\newacronym{EMT}{EMT}{energy-momentum tensor}

\newcommand{\RHheaderline}{\textsf{TTK-25-49, P3H-25-116, December 2025}}
\fancypagestyle{firstpage}
{
  
  \fancyhead[R]{\RHheaderline}
}
\title{Quark-mass effects in gradient-flow observables
  through next-to-next-to-leading order in QCD}
\author{R.V.~Harlander and R.H.~Mason}
\affil{TTK, RWTH~Aachen University, Sommerfeldstr.~16, 52074~Aachen, Germany}
\date{}

\begin{document}

\maketitle
\thispagestyle{firstpage}

\begin{abstract}
  We provide results for the vacuum expectation values of the flowed action
  density, the quark condensate, and the quark kinetic operator in the
  gradient-flow formalism. We work in $\nf$-flavor QCD, keeping the heaviest
  quark massive and all others massless. The vacuum expectation values of
  these operators are calculated numerically through next-to-next-to-leading
  order QCD, providing important input for the extraction of fundamental QCD
  parameters from lattice calculations. While the focus is on charm- and
  bottom-quark mass effects, we provide the results in a form that is
  independent of the specific quark mass.
\end{abstract}
\tableofcontents
\glsresetall 



\section{Introduction}\label{sec:intro}

The \gff~\cite{Narayanan:2006rf,Luscher:2009eq,Luscher:2010iy,
  Luscher:2011bx,Luscher:2013cpa} is a method to exponentially suppress the
high-momentum modes of fields in a symmetry-preserving way. This suppression
is governed by the flow time $t$, which is a parameter of mass dimension~\two.
Flowed fields are then defined to exist in the \five-dimensional space
$(x,t\geq 0)$, such that they regain their interpretation as physical fields
on the boundary at $t=0$. At $t>0$, they are determined by the flow equations,
which are differential equations of first-order in $t$.

The \gff\ was first developed for use in lattice field theory and has since
seen much utility on the lattice, predominantly for the purpose of
scale-setting~\cite{Luscher:2010iy,BMW:2012hcm,
  FlavourLatticeAveragingGroupFLAG:2024oxs}.  While on the lattice, the flow
equations are solved numerically, the gradient flow can also be approached
perturbatively~\cite{Luscher:2011bx,Luscher:2013cpa}. In fact, in some
applications, the combination of lattice and perturbative results is crucial
within the \gff. For example, the \sftx\ allows one to relate flowed and
regular operators via perturbative matching coefficients. This circumvents the
conventional operator renormalization on the lattice and thus avoids operator
mixing when taking the continuum limit.  In a sense, the \gff\ offers a
renormalization scheme which is both accessible on the lattice and in
perturbation theory. Examples for practical applications are the computation
of meson lifetimes~\cite{Suzuki:2020zue,Black:2023vju}, the energy-momentum
tensor~\cite{Suzuki:2013gza,Taniguchi:2020mgg}, and moments of parton-density
functions~\cite{Shindler:2023xpd,Francis:2025rya}.

The main motivation for the current paper, on the other hand, is the
determination of the strong coupling (or, equivalently, the \qcd\ scale
$\Lambda_\text{\qcd}$) and the quark masses. Being fundamental parameters of
\qcd, any uncertainty in their value directly feeds into the computation of
many processes and consequently much importance must be attached to
quantifying them as precisely as possible. Currently, there are numerous
efforts devoted to computing these quantities (for reviews, see
\citeres{FlavourLatticeAveragingGroupFLAG:2024oxs,
  ParticleDataGroup:2024cfk}), some of them based on the \gff~(see, e.g.,
\citeres{Hasenfratz:2023bok,Wong:2023jvr,Schierholz:2024lge,
  Larsen:2025wvg,Brida:2025gii, Takaura:2025pao,Black:2025gft}). The \vevs{}
of certain quark bilinear operators and the gluonic action density play a
crucial role in many of these methods. In a suitable renormalization scheme,
these are \rg\ invariant quantities and can be computed both on the lattice
and in perturbation theory. Comparison of the two thus provides a link between
the hadronic phase of \qcd\ and its fundamental Lagrangian parameters.

The accuracy of such extractions is determined by both the uncertainty of the
lattice and the perturbative calculation, of course. As a general rule, in a
perturbative \qcd\ calculation, the residual dependence of a fixed-order
result on the renormalization scale $\mu$ is used as an estimate of the
theoretical uncertainty of that result. At \lo, the $\mu$-dependence is
typically monotonous, making it difficult to choose a particular interval for
the scale variation. Since the quark bilinear operators are independent of the
strong coupling, the scale variation even vanishes at \lo\ if the \vevs{} are
expressed in terms of on-shell quark masses, while at \nlo, it is
monotonous~\cite{Takaura:2025pao}. In the case of the action density, the
quark-mass dependence of its \vev\ occurs only at \nlo, while from the
massless calculation we know that \nnlo\ effects are quite crucial for a
precise perturbative prediction. These observations motivate a calculation of
the quark-mass effects for the corresponding \vevs{} to \nnlo, which is the
purpose of this paper.

Its remaining structure is as follows. In \cref{sec:FlowEquations} we cover
the basics of the \gff\ within perturbation theory. \cref{sec:vevs} defines
the observables under consideration and their renormalization; it also briefly
discusses the methods we employ to compute them through \nnlo\ \qcd.  In
\cref{sec:Results} we provide numerical results in a form that is independent
of the specific quark mass. These results are also provided in
computer-readable format in an ancillary file accompanying this
paper. \cref{sec:pheno} then studies the mass effects in the context of
physical parameters, and \cref{sec:conclusions} contains our conclusions.


\section{Flow equations} \label{sec:FlowEquations}

Working in Euclidean space, we write the quantized regular \qcd\ Lagrangian as
\begin{equation}\label{eq:Flow:axes}
  \begin{aligned}
    \mathcal{L}_\text{\qcd} &= \frac{1}{4g_\bare^2}F^a_{\mu\nu}F^a_{\mu\nu}
    +\sum^{\nf}_{f=1}\bar{\psi}_f\left( \slashed{D}^F+m_{f,\bare}\right)\psi_f
    + \mathcal{L}_\text{gf} + \mathcal{L}_\text{gh}\,,
  \end{aligned}
\end{equation}
with the gauge-fixing and ghost terms
\begin{equation}\label{eq:Flow:jasp}
  \begin{aligned}
  \mathcal{L}_\text{gf}&=\frac{1}{2 \xi_\bare}\left( \partial_\mu A^a_\mu
  \right)^2\,,&&& \mathcal{L}_\text{gh}&= \partial_\mu \bar{c}^a D^{ab}_\mu
  c^b\,,
  \end{aligned}
\end{equation}
respectively. Here, $A_\mu^a$ is the gluon field with color index $a$,
$\psi_f$ is the Dirac spinor for a quark of flavor $f$ and bare mass
$m_{f,\bare}$, $c$ and $\bar{c}$ are the ghost and anti-ghost fields,
respectively, $\xi_\bare$ is the bare gauge parameter, and $g_\bare$ the bare
gauge coupling.  The field strength tensor is defined as
\begin{equation}\label{eq:Flow:juda}
  \begin{aligned}
    F^a_{\mu\nu}&=\partial_\mu A^a_\nu - \partial_\nu A^a_\mu +f^{abc}A^b_\mu
    A^c_\nu\,,
  \end{aligned}
\end{equation}
with the SU(3) structure constants $f^{abc}$, and the covariant derivative in
the fundamental representation is
\begin{equation}\label{eq:Flow:kahn}
  \begin{aligned}
    D_\mu^F &= \partial_\mu + A_\mu^a \frac{\lambda^a}{2}\,,
  \end{aligned}
\end{equation}
with $\lambda^a$ the Gell-Mann matrices.

The \gff\ can be defined by adding the following two terms to the regular
\qcd\ Lagrangian:
\begin{equation}\label{eq:Flow:arad}
  \begin{aligned}
    \mathcal{L}^\text{flow}_\text{\qcd} &= \mathcal{L}_\text{\qcd} +
    \mathcal{L}_B + \mathcal{L}_\chi\,,
  \end{aligned}
\end{equation}
with
\begin{equation}\label{eq:Flow:adel}
  \begin{aligned}
    \mathcal{L}_B &= \int_0^\infty\dd t\, L^a_\mu\mathcal{F}_\mu^a\,,&&&
    \mathcal{L}_\chi &=
    \sum_{f=1}^{\nf}\int_0^\infty\dd t\,\left[ \bar{\lambda}_f\mathcal{F}^f_{\chi}
    + \bar{\mathcal{F}}^f_{\chi}\lambda_f\right]\,,\\
    \mathcal{F}_{B,\mu}^a &= \partial_t B_\mu^a
    - \mathcal{D}^{ab}_\nu G_{\nu\mu}^b
    - \kappa \mathcal{D}^{ab}_\mu\partial_\nu B_\nu^b\,,&&&
    \mathcal{F}^f_{\chi} &= 
        \left(\partial_t - \Delta
    - \kappa\, \partial_\nu B_\nu^b\right)\chi_f\,,\\
    &&&&\bar{\mathcal{F}}^f_{\chi} &= 
        \bar{\chi}_f\left(\overleftarrow{\partial}_t - \overleftarrow{\Delta}
    - \kappa\, \partial_\nu B_\nu^b\right)\,.
  \end{aligned}
\end{equation}
The flowed covariant derivative in the adjoint representation is
defined as
\begin{eqnarray}
  \mathcal{D}^{ab}_\mu&=&\delta^{ab}\partial_\mu -f^{abc}B^c_\mu\,,
\end{eqnarray}
and $\Delta$ and $\overleftarrow{\Delta}$ are the flowed covariant d'Alembert
operator and its left-acting counterpart, defined according to
\begin{equation}\label{eq:Flow:faro}
  \begin{aligned}
  \Delta&=\mathcal{D}^{F}_\mu\mathcal{D}^{F}_\mu\,,&&&
  \overleftarrow{\Delta}&=\overleftarrow{\mathcal{D}}^{F}_\mu
  \overleftarrow{\mathcal{D}}^{F}_\mu\,,\\
  \mathcal{D}_\mu^{F} &= \partial_\mu + B_\mu^a\frac{\lambda^a}{2}\,,&&&
  \overleftarrow{\mathcal{D}}_\mu^{F} &= \overleftarrow{\partial}_\mu
  - B_\mu^a\frac{\lambda^a}{2}\,,
  \end{aligned}
\end{equation}
The flowed field strength tensor $G_{\mu\nu}^a$ is related to
\cref{eq:Flow:juda} by replacing $A^a_\mu$ by $B^a_\mu$.

The Euler-Lagrange equations for the Lagrange multiplier fields $L_\mu^a$,
$\lambda_f$, and $\bar\lambda_f$ yield the actual flow equations for the
flowed gluon field $B^a_\mu$ and the flowed quark fields $\chi_f$ and
$\bar{\chi}_f$:
\begin{equation}\label{eq:Flow:ahom}
  \begin{aligned}
    \mathcal{F}_{B,\mu}^a &= 0\,,&&&
    \mathcal{F}^f_\chi &= 0\,,&&&
    \bar{\mathcal{F}}^f_\chi &= 0\,.
  \end{aligned}
\end{equation}
The flowed fields thus depend on the flow-time $t$ which is a new parameter
with mass dimension $=-2$. In \cref{eq:Flow:adel}, $\kappa$ is an unphysical
gauge parameter associated with the gradient flow and is generally taken to
unity.

In addition to \cref{eq:Flow:ahom}, one imposes the ``initial conditions''
\begin{equation}\label{eq:Flow:john}
  \begin{aligned}
    B_\mu(t=0,x)&=A_\mu(x)\,,&&&\chi_f(t=0,x)&=\psi_f(x)\,,
    &&&\bar{\chi}_f(t=0,x)&=\bar{\psi}_f(x)\,.
  \end{aligned}
\end{equation}
\cref{eq:Flow:arad,eq:Flow:john} can be translated into Feynman rules
involving simple generalizations of the regular quark and gluon propagators,
as well as a gluon and a quark \textit{flowline}, induced by the mixing with
their respective Lagrange multiplier fields. As opposed to the propagators,
the flowline does not involve a momentum pole. Furthermore, in addition to the
regular \qcd\ vertices, the flow terms of \cref{eq:Flow:adel} induce a number
of vertices between flowed fields and their Lagrange multiplier fields. The
complete set of Feynman rules for flowed \qcd\ can be found in
\citere{Artz:2019bpr}.


\section{Vacuum expectation values}\label{sec:vevs}



\begin{figure}
  \begin{center}
    \begin{tabular}{cc}
      \includegraphics[width=3cm]{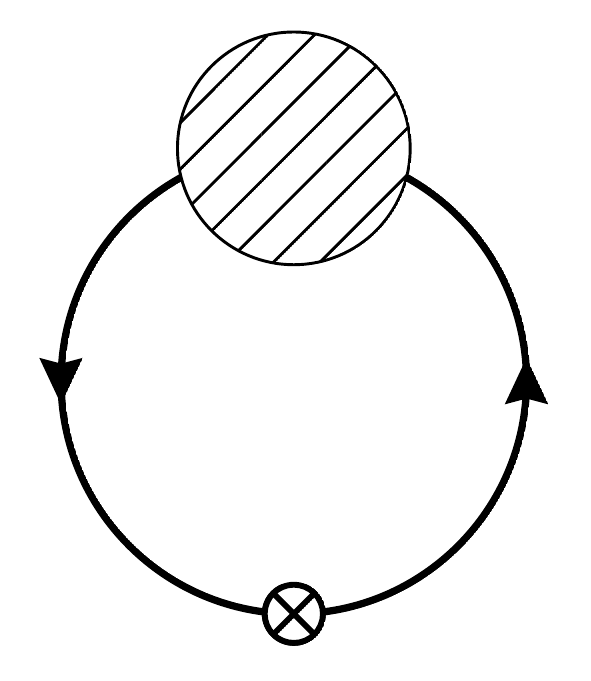}  &
      \includegraphics[width=3cm]{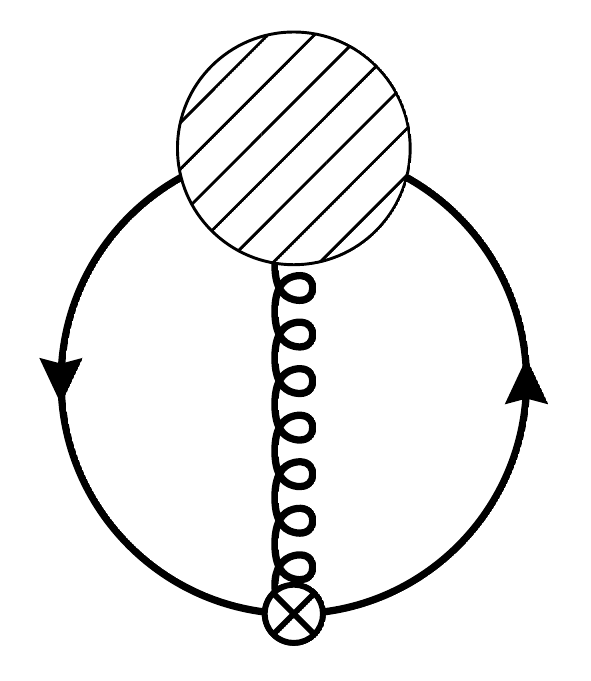}  \\
      (a)&(b)\\
    \end{tabular}
    \begin{tabular}{ccc}
      \includegraphics[width=3cm]{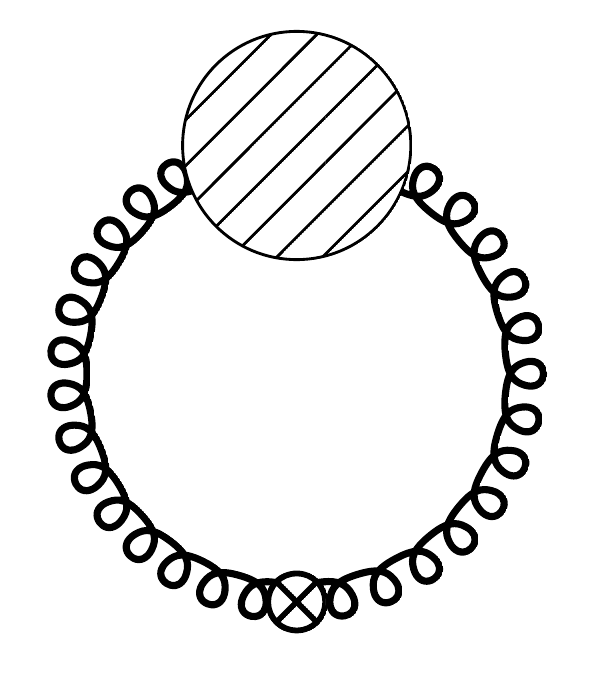} &
      \includegraphics[width=3cm]{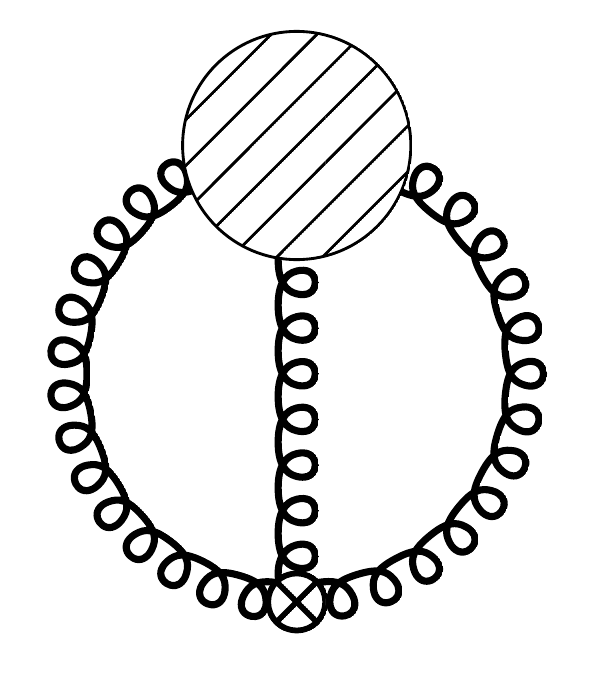}
      &
      \includegraphics[width=3cm]{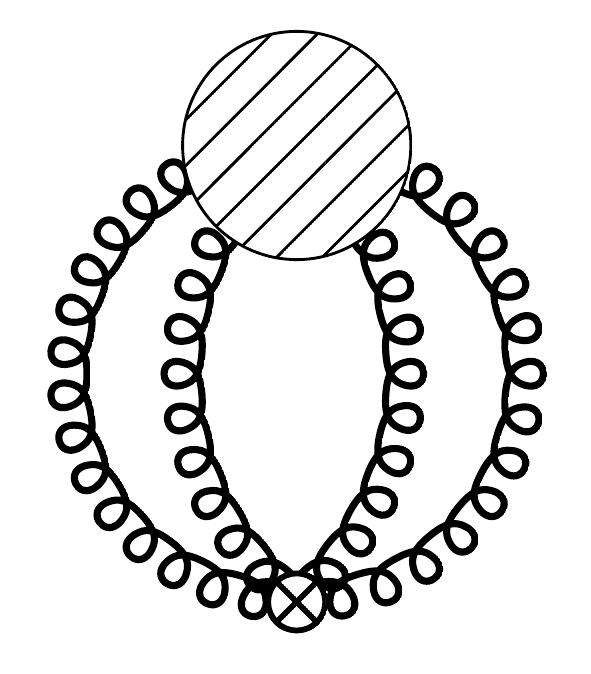}  \\
      (c) & (d) & (e)
    \end{tabular}
  \end{center}
  \caption{Diagrams contributing to the vacuum expectation value of the quark
    condensate $S_f(t)$ (a), the quark kinetic operator~$R_f(t)$ (a/b), and
    the gluon condensate $E(t)$ (c--e). The hatched bubble represents an
    arbitrary flowed-\qcd\ sub-diagram. Some of the lines can also be
    flowlines. All Feynman diagrams in this paper were created with
    \texttt{FeynGame}~\cite{Bundgen:2025utt,Harlander:2024qbn,
      Harlander:2020cyh}.
    \label{fig:GenFeynmanDiagrams}}
\end{figure}


\subsection{Definitions}


\begin{figure}
  \begin{center}
    \begin{tabular}{cc}
      \includegraphics[width=3cm]{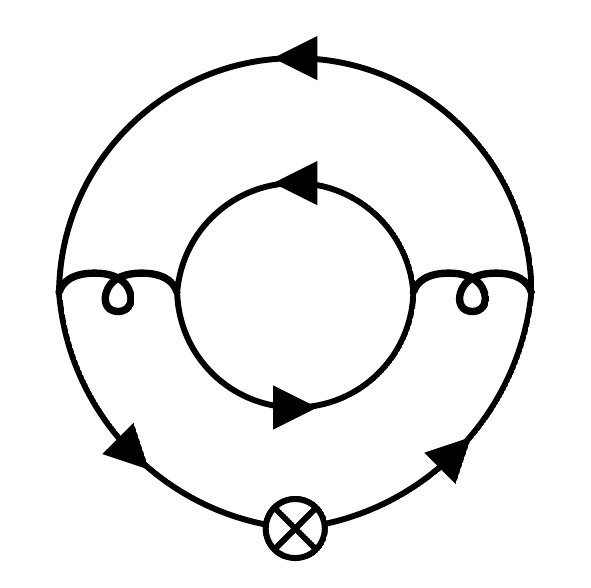}   &
      \includegraphics[width=3cm]{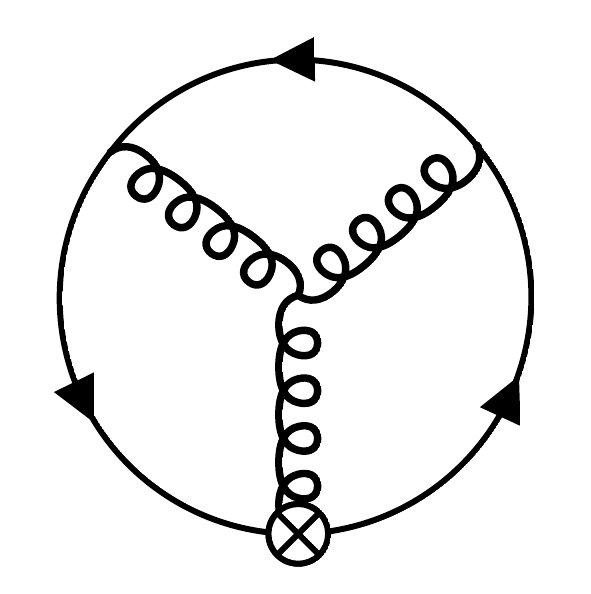}  \\
      (a)&(b)\\
    \end{tabular}
    \begin{tabular}{ccc}
      \includegraphics[width=3cm]{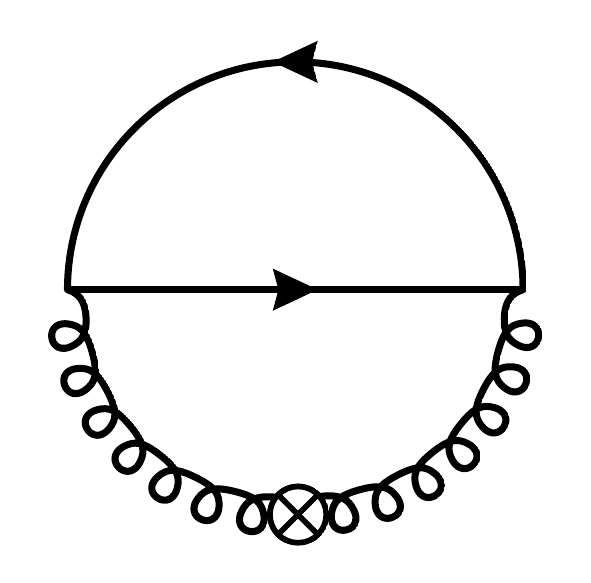} &
      \includegraphics[width=3cm]{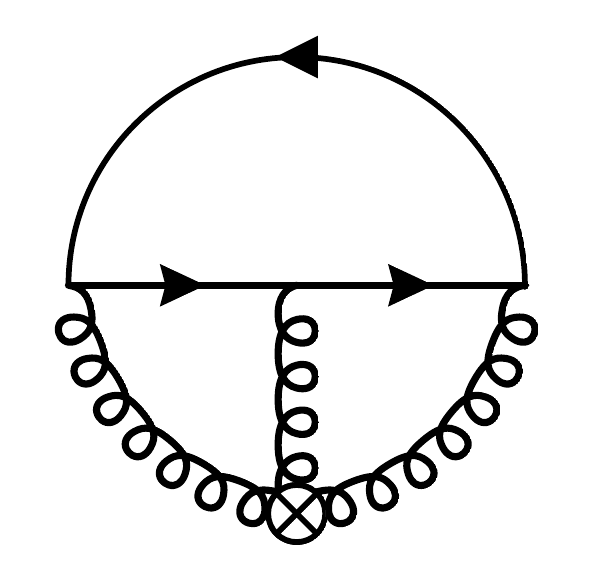}
      &
      \includegraphics[width=3cm]{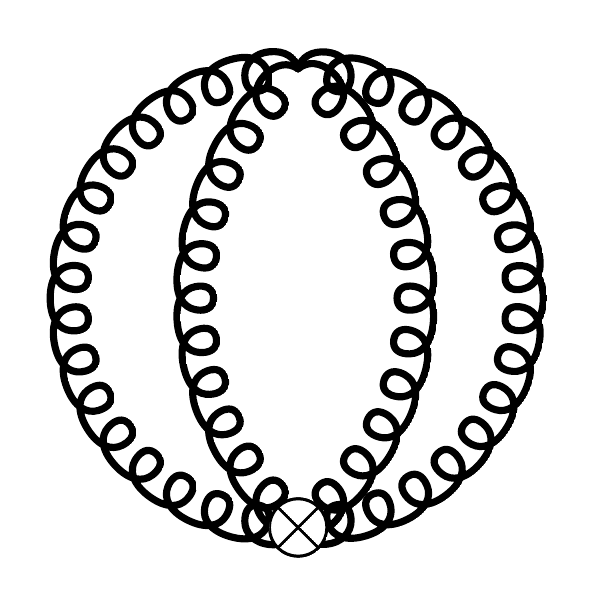}   \\
      (c) & (d) & (e)
    \end{tabular}
  \end{center}
  \caption{Specific diagrams contributing to $S(t)$ (a), $R(t)$ (a/b) at
    the \three-loop level and $E(t)$ at the \two- (c) and \three-loop level
    (d) and (e). \label{fig:SpecFeynmanDiagrams}}
\end{figure}


For the purposes of this study we will consider the mass effects on the
\vevs{} of three operators: the quark condensate $S_f(t)$, the quark kinetic
density $R_f(t$), and the action density $E(t)$. These are described by
\begin{equation}\label{eq:Flow:bank}
  \begin{aligned}
  S_f(t) &= \langle \bar{\chi}_f(t)\chi_f(t)
  \rangle,\\
  R_f(t) &= \langle
  \bar{\chi}_f(t)\overleftrightarrow{\slashed{\mathcal{D}}}\chi_f(t)
  \rangle\,,
  \\
  E(t) &= \frac{1}{4}\langle
  G^a_{\mu\nu}(t)G^{a,\mu\nu}(t) \rangle\,,
  \end{aligned}
\end{equation}
where $\overleftrightarrow{\mathcal{D}_\mu}
=\mathcal{D}^{F}_\mu-\overleftarrow{\mathcal{D}}^{F}_\mu$.  Diagrammatically,
these are represented by vacuum diagrams with an operator
insertion. \cref{fig:GenFeynmanDiagrams} displays the general form of diagrams
for each of these processes, with \cref{fig:SpecFeynmanDiagrams} providing
specific examples in each case.

The quark condensate $S_f(t)$ vanishes in the massless-quark case due to
chiral symmetry. For an arbitrary quark mass, it was first calculated at
\one-loop level in \citeres{Luscher:2013cpa,Harlander:2021esn}, and to two
loops in \citere{Takaura:2024ejs}. Keeping only the leading (linear) term in
the quark masses, the \three-loop result was obtained in
\citere{Artz:2019bpr}.  For the massless case, $R_f(t)$ is known up to the
\three-loop level~\cite{Makino:2014taa,Harlander:2018zpi,Artz:2019bpr}, while
mass effects were evaluated to two loops in \citere{Takaura:2025pao}. $E(t)$
only receives massive contributions at two loops and above, which have been
considered in \citere{Harlander:2016vzb}, with the massless case being taken
to three loops~\cite{Luscher:2010iy,Harlander:2016vzb}. At \two-loop level,
only a single diagram contributes to the mass effects of $E(t)$, shown in
\cref{fig:SpecFeynmanDiagrams}\,(c).


\subsection{Renormalization}\label{sec:Renormalization}

Renormalization in the \gff\ is carried out in the usual way, with the
renormalized fields and parameters being related to their bare counter parts
by
\begin{equation}\label{eq:Reno:duad}
  \begin{aligned}
        B_\mu &= \sqrt{Z_B} B_{0\mu}\,,
 &&&
    \alpha_s &= \left(\frac{\mu^2
    e^{\EulerGamma}}{4\pi} \right)^\ep Z_\alpha
  \frac{g_0^2}{4\pi}\,,\\
  \chi &= \sqrt{Z_\chi} \chi_0\,,
  &&&
   m_f &= Z_m m_{f0}\,,\\
   \bar{\chi} &= \sqrt{Z_\chi} \bar{\chi}_0\,,
  \end{aligned}
\end{equation}
where the renormalization constants are Laurent series in the dimensional
regularisation parameter $\ep=(4-D)/2$, with $D$ being the number of
space-time dimensions which we adopt for the loop-momentum integration. As the
flow time acts as a \uv\ cutoff for the high momentum modes, additional
operator renormalization is not required for flowed composite
operators~\cite{Luscher:2011bx}.

The coupling and mass renormalization constants are the same as in regular
\qcd. In the \msbar\ scheme, they read
\begin{equation}\label{eq:Reno:kino}
  \begin{aligned}
    Z_\alpha &= 1-\api\frac{\beta_0}{2\ep} +
    \api^2\left(\frac{3\beta_0^2}{8\ep^2} - \frac{\beta_1}{4\ep}\right) +
    \order{\api^3}\,,\\
    Z_m^\text{\msbar} &= 1 - \api\,\frac{\gamma^m_0}{\epsilon}  +
    \api^2\left[\frac{\gamma^m_0}{2\epsilon^2}\left(\gamma^m_{0}  +
      \beta_0\right)
      -\frac{\gamma^m_{1}}{2\epsilon}\right]  +
    \mathcal{O}(\api^3)\,,
  \end{aligned}
\end{equation}
with
\begin{equation}\label{eq:Resu:anew}
  \begin{aligned}
    \api(\mu) &= \frac{\alpha_s(\mu)}{4\pi}\,,
  \end{aligned}
\end{equation}
and
\begin{equation}\label{eq:Resu:byng}
  \begin{aligned}
    \beta_0 &= 11 - \frac{2}{3}\nf\,,&&&
    \beta_1 &= 102 - \frac{38}{3}\nf\,,\\
    \gamma^m_0 &= 4\,,&&&
    \gamma^m_1 &= \frac{202}{3} - \frac{20}{9}\nf\,.
  \end{aligned}
\end{equation}
As will become clear below, it is convenient to present our results with the
quark mass renormalized in the on-shell scheme. In this case, the mass
renormalization constant reads
\begin{equation}\label{eq:Reno:dyad}
  \begin{aligned}
    Z_m^\text{\abbrev{OS}} &= c_mZ_m^\text{\msbar}\,,
  \end{aligned}
\end{equation}
with
\begin{equation}\label{eq:Reno:krim}
  \begin{aligned}
    c_m &= 1
    + \api\delta_m^{(1)}
    + \api^2\delta_m^{(2)} + \order{\api^3}\,,
  \end{aligned}
\end{equation}
and
\begin{equation}\label{eq:Reno:floc}
  \begin{aligned}
    \delta_m^{(1)} &= \frac{16}{3} + 4\ln\frac{\mu^2}{\bar{m}^2(\mu)}\,,\\
    \delta_m^{(2)} &= 215.094- 16.662 \nl +\left( 109.556-5.778 \nl\right)\ln\frac{\mu^2}{\bar{m}^2(\mu)}\\
    &~~~+\left(28.667 -1.333 \nl\right)\ln^2\frac{\mu^2}{\bar{m}^2(\mu)}  \,.
  \end{aligned}
\end{equation}
where $\nl$ is the number of massless quarks. The full analytic expressions
are available in \citere{Chetyrkin:1999qi}.  Since we evaluate the integrals
numerically, we need to insert the mass counter terms at the integrand level
and re-expand the corresponding propagators in the coupling constant as
follows:
\begin{eqnarray}\label{eq:ren:massexp}
\frac{1}{p^2+m_0^2}&=&\frac{1}{p^2+m^2}\left(1-\frac{\delta m^2}{p^2+m^2}
+\mathcal{O}\left(\delta m^2\right) \right),\nonumber
\end{eqnarray} 
where $\delta m^2 = m^2(Z_m^2-1)=\mathcal{O}(\alpha_s)$. This means that we
need to evaluate integrals with higher powers of the propagators, albeit only
at lower loop orders, because each additional power is accompanied by a factor
$\alpha_s$. Such integrals can be calculated in the same way as described in
\cref{sec:ProcessesComputation} below.

While the flowed-gluon field renormalization can be set
to~\cite{Luscher:2010iy,Luscher:2011bx}
\begin{equation}\label{eq:Reno:east}
  \begin{aligned}
    Z_B &= 1\,,
  \end{aligned}
\end{equation}
the flowed-fermion field renormalization in the \msbar\ scheme
reads~\cite{Harlander:2018zpi,Artz:2019bpr}
\begin{eqnarray}
Z^\text{\msbar}_\chi &=&1+\api\left[\frac{3
  \ccf}{\epsilon}\right]+\api^2
\frac{\ccf}{2\epsilon}\left[\frac{3}{\epsilon}\left(3\ccf
  - \frac{11}{3}\cca-\frac{4}{3}\ctr \nf\right)\right.\nonumber\\
  &&~~~\left.+\cca\left(\frac{223}{6}-8\ln2\right)
  -\ccf\left(\frac{3}{2}+8\ln2\right)
  -\frac{22}{3}\ctr \nf \right]+\mathcal{O}(\api^3)\,,
\end{eqnarray}
where $\EulerGamma\approx 0.5772$ is the Euler-Mascheroni constant, $\ccf$,
$\cca$, and $\ctr$ are the Casimir factors and the trace normalization of the
gauge group; in \qcd, it is
\begin{equation}\label{eq:Reno:krti}
  \begin{aligned}
\ccf=\frac{4}{3}\,,\quad \cca=3\,,\quad \text{and}\quad \ctr=\frac{1}{2}\,.
  \end{aligned}
\end{equation}

Since the $\msbar$ scheme is intrinsically perturbative and we would like to
be able to combine our results with the corresponding lattice results, it is
preferable to adopt a non-minimal, regularization-independent renormalization
scheme for the fermions which leads to \rg\ invariant quantities. One option
is the so-called ringed scheme, defined as
\begin{equation}\label{eq:Reno:hern}
  \begin{aligned}
    \mathring{Z}_\chi\sum_f R_f(t)\Big|_{m=0} \equiv -\frac{2\nc\nf}{(4\pi
      t)^2}\,,
  \end{aligned}
\end{equation}
with $\nc=3$ the number of colors, $R_f(t)$ from \cref{eq:Flow:bank}, and
\begin{equation}\label{eq:Reno:judy}
  \begin{aligned}
    \mathring{Z}_\chi &= \zeta_\chi\,Z^\text{\msbar}_\chi\,.
  \end{aligned}
\end{equation}
Through \nnlo, the conversion factor
reads~\cite{Harlander:2018zpi,Artz:2019bpr}
\begin{equation}\label{eq:Reno:jube}
  \begin{aligned}
    \zeta_\chi &= 1+\api\left(3\, \ccf \lmut-3\,\ccf \ln 3 -4
    \,\ccf \ln 2 \right) \\
    &~~~+\api^2 \Bigg\{\frac{\ccf}{6}\Big[
      \cca\Big( - 136\,\ln 2 - 66\,\ln 3
      + 223\Big) + \ccf\Big( - 120\ln 2 \\
      &~~~~~~ - 54\ln 3 - 9\Big)
      + \ctr\nf\Big(32\ln 2
      + 24 \ln 3 - 44\Big)\Big]\lmut^2 \\
    &~~~~~~  +\frac{\ccf}{2} \Big(11\,\cca + 9\,\ccf - 4\,\ctr\nf\Big)\lmut 
    -23.7947\, \cca \ccf + 30.3914\, \ccf^2\\
    &~~~~~~ -3.9226\, \ccf\ctr\nf \Bigg\}.\\
  \end{aligned}
\end{equation}
The renormalized quantities
\begin{eqnarray}
  \mathring{S}_f(t) = \mathring{Z}_\chi S_{f}(t)~~~\text{and}~~~
  \mathring{R}_f(t)=\mathring{Z}_\chi R_{f}(t)
\end{eqnarray}
are then \rg\ invariant. These two quantities, together with the \rg-invariant
\vev\ of the action density, $E(t)$, are the main focus of this paper.  Of
course, they also allow one to evaluate the \rg-invariant ratios
\begin{equation}\label{eq:Reno:dole}
  \begin{aligned}
  r_{f,a}(t) &= \frac{S_{f}(t)}{R_{f}(t)}\,,\\
  r_{f,b}(t)&=\frac{R_{f}(t)\qquad}{R_{f}(t)|_{m=0}} =
  - \frac{(4\pi t)^2}{2\nc}\mathring{R}_f(t)\,,\\
  r_{f,c}(t)&=m_f\frac{d}{dm_f}r_{f,a}(t)
  \end{aligned}
\end{equation}
suggested in \citere{Takaura:2025pao} for the quark-mass determination.

The bulk of our results is obtained under the assumption that there is one
massive quark and $\nl=\nf-1$ light quarks, where we focus on $\nf=4$ and
$\nf=5$. When calculating $S_f$ and $R_f$, we assume $f$ to be the massive
quark. The effect of a second massive quark will be briefly discussed at the end
of \cref{sec:pheno}.


\subsection{Computation}\label{sec:ProcessesComputation}


\begin{table}[h]
  \begin{center}
    \caption{Number of integrals involving zero, one, and two masses, and the
      number of Feynman diagrams, split into \lo, \nlo, and \nnlo\ for $
      S_f(t)$, $ R_f(t)$ and $ E(t)$.
    \label{tab:IntTable}}
    \begin{tabular}{l|ccc|ccc|ccc}
      &\multicolumn{3}{c|}{ $S_f(t)$} &
      \multicolumn{3}{|c|}{ $R_f(t)$} & \multicolumn{3}{|c}{ $E(t)$}\\
      \cline{2-10}
      &\lo & \nlo & \nnlo & \lo & \nlo & \nnlo & \lo & \nlo & \nnlo \\\hline
      zero masses & 0 & 2 & 177  & 3 & 11 & 790   & 1 & 23 & 1990 \\
      one mass & 3 & 30 & 978 & 1 & 47 & 1922  & 0 & 10 & 409 \\
      two masses& 0 & 0 & 71  & 0 & 0  & 182   & 0 & 0 & 20 \\
      \hline
      diagrams& 1 & 8 & 258 & 1 & 11 & 377 & 1 & 17 & 419\\
    \end{tabular}
  \end{center}
\end{table}


As opposed to the \nlo\ calculation~\cite{Luscher:2010iy,Hieda:2016lly,
  Harlander:2016vzb,Takaura:2025pao}, the increased complexity at \nnlo\ makes
automation indispensible.  \cref{tab:IntTable} provides a breakdown of the
number of diagrams and integrals that appear in the computation of each of the
\vevs{}.\footnote{These numbers are supplied without any attempt to reduce
them via integration-by-parts relations.}  While this list provides an
approximate view of the comparative computation cost in calculating each
process, it does not reflect the increase in complexity by the inclusion of
non-zero quark masses.

A general overview of the setup used in the computations in this paper can be
found in \citere{Artz:2019bpr}, but in brief, computations proceed as in
regular perturbation theory. We generate the necessary Feynman diagrams using
\verb|qgraf|~\cite{Nogueira:1991ex,Nogueira:2006pq} and visualize them with
the help of \verb|FeynGame3|~\cite{Bundgen:2025utt}; the Feynman rules are
inserted using \verb|tapir|~\cite{Gerlach:2022qnc} and
\verb|exp|~\cite{Harlander:1998cmq} such that the Feynman integrals can be
produced in the symbolic manipulation language
\verb|FORM|~\cite{Vermaseren:2000nd,Kuipers:2012rf}; the integrals are then
evaluated.  Due to the addition of an exponential smoothing term in the
computation of processes involving flowed fields, the integrals to be
evaluated are distinct from standard Feynman integrals. They are also highly
structured though, allowing for integral reduction via integration-by-parts
relations~\cite{Chetyrkin:1981qh,Artz:2019bpr};
\verb|Kira+FireFly|~\cite{Maierhofer:2017gsa,Maierhofer:2018gpa,
  Klappert:2019emp,Klappert:2020nbg} can be used for this purpose in
principle. In our case, we find that the reduction is less efficient than a
direct numerical evaluation of the integrals.

To be specific, the scalar integrals we encounter at $l$-loop level, for
$l\leq 3$, are of the general form
\begin{equation}\label{eq:vevs:gong}
  \begin{aligned}
 &f({\bf c}, { \bf
    b},\{\{n_1,m_1\},\ldots,\{n_n,m_n\},\{n_{1,2},m_{1,2}\},\ldots,
  \{n_{n-1,n},m_{n-1,n}\}\})=\\ &\qquad=(4\pi
  t)^{\frac{ld}{2}} \int^1_0 {\bf u}^{\bf c} d{\bf u} \prod_{i=1}^n \int_{k_i}
   \frac{\exp(-t b_i k_i^2)}{(t(k_i^2+m_i^2))^{n_i} }
  \prod_{j>i}^n \frac{\exp(-t b_{i,j}
    k_{i,j}^2)}{(t(k_{i,j}^2+m_{i,j}^2))^{n_{i,j}} }\,,
  \end{aligned}
\end{equation}
where $k^2_{i,j}=(k_i-k_j)^2$ with $b_{i,j}$, $n_{i,j}$ and $m_{i,j}$ being
parameters associated with these momenta. Note that factors of the flow time
are included to ensure the term is dimensionless, and the ${\bf b}$ parameters
can be functions of the ${\bf u}$ parameters. The latter arise from
contributions due to flowlines.

While the integrals in \cref{eq:vevs:gong} can be evaluated analytically for
$l=1$, the exact solution at two loops is known only for special values of the
parameters, in particular for the massless case
$m_i=m_{i,j}=0$~\cite{Harlander:2018zpi}. At \three-loop level, even this case
has not been fully solved. We therefore resolve to a numerical evaluation of
the integrals.

A program designed specifically for this purpose is
\verb|ftint|~\cite{Harlander:2024vmn}. It provides an interface to
\verb|pySecDec|~\cite{Borowka:2017idc} which applies sector decomposition and
subsequent numerical integration of the integrals. The result is a Laurent
series in $\ep$ with numerical coefficients. The cancellation of the pole
terms in $\ep$ in the renormalized results to the numerical accuracy of the
computation provides an obvious though non-trivial check of our results.

The sector decomposition of the integrals is independent of the specific mass
values $m_i$ and $m_{i,j}$. \verb|ftint| allows one to separate this step from
the subsequent numerical integration, which can then be performed for
arbitrary values of the masses.  This is particularly important since in our
case the sector decomposition is typically more computationally expensive than
the numerical integration.



\section{Results}\label{sec:Results}


The main results of this paper constitute the numerically evaluated component
values of the perturbative series of the \vevs{} defined in
\cref{eq:Flow:bank}. Since we provide only numerical results for these
quantities, we need to define an adequate range for the flow time. Its value
should be distant from the limiting scales, given by the square of the lattice
spacing $a$ and size $L$, as well as $1/\Lambda^2_{\text{\qcd}}$ which is
associated with the breakdown of perturbation theory. Following the conditions
on the flow time $t$ suggested by \citere{Takaura:2024ejs,Takaura:2025pao},
\begin{equation}
  \begin{aligned}
  a^2\ll &\ 8t\ll L^2\,,&&& 8t\ll &\ \Lambda^{-2}_{\text{\qcd}}\,,\\
  0.1\ll &\ 8\bar{m}_c^2 t\ll 20\,,&&&
  1.0\ll &\ 8\bar{m}_b^2 t\ll 200\,,\label{eq:trange}    
  \end{aligned}
\end{equation}
a set of 200~points considered between $0.001\leq m^2t\leq 63$ should be
sufficient to cover our region of interest for both the case of the charm and
the bottom quark.  Note, often linear interpolation is used between points on
this grid.  In the following we will present these data in a form that is
independent of the specific massive quark flavor under consideration.


\subsection{The scalar quark density}\label{sec:sres}

Up to the perturbative order $N$, the \vev\ of the scalar quark density in the
ringed scheme can be written as
\begin{equation}\label{eq:Resu:jael}
  \begin{aligned}
    \mathring{S}_f(t) &= -\frac{\nc m_f}{8\pi^2 t}\,\sum_{n=0}^N
  \api^n(\mu) s_{n}(m_f,t,\mu)\,,&&&
    s_n(m,t,\mu) &= \sum_{k=0}^n s_{n,k}(m^2t)\lmut^k\,,
  \end{aligned}
\end{equation}
where
\begin{equation}\label{eq:Resu:anew1}
  \begin{aligned}
    \lmut = \ln 2\mu^2t + \EulerGamma\,.
  \end{aligned}
\end{equation}
The coefficients $s_{n,k}(m^2t)$ are dimensionless. At \nnlo, we further split
them into
\begin{equation}\label{eq:Resu:epic}
  \begin{aligned}
    s_{2,k} &= s_{2,k,0} + \nl s_{2,k,1}\,,
  \end{aligned}
\end{equation}
where $\nl$ is the number of light (massless) quarks.

For simplicity of the presentation, let us assume for the moment that the mass
$m$ is renormalized in the on-shell scheme. We will provide the required
conversion to the \msbar\ scheme further below. Due to the \rg\ invariance of
$\mathring{S}_f$ and the fact that it starts at order $\alpha_s^0$, a
logarithmic term only appears at \nnlo. It is given by the non-logarithmic
\nlo\ coefficient as
\begin{equation}\label{eq:Resu:hook}
  \begin{aligned}
    s_{2,1} = \beta_0\,s_{1,0}\,,
  \end{aligned}
\end{equation}
where $\beta_0$ is given in \cref{eq:Resu:byng}.  The \lo\ result can be
obtained analytically~\cite{Luscher:2013cpa,Harlander:2021esn}:
\begin{equation}\label{eq:Resu:fief}
  \begin{aligned}
    s_{0,0}(z) &= 1 - 2z\, e^{2z}\,\Gamma(0,2z)\,.
  \end{aligned}
\end{equation}


\begin{figure}[!ht]
    \centering
    \includegraphics[width=0.9\textwidth]{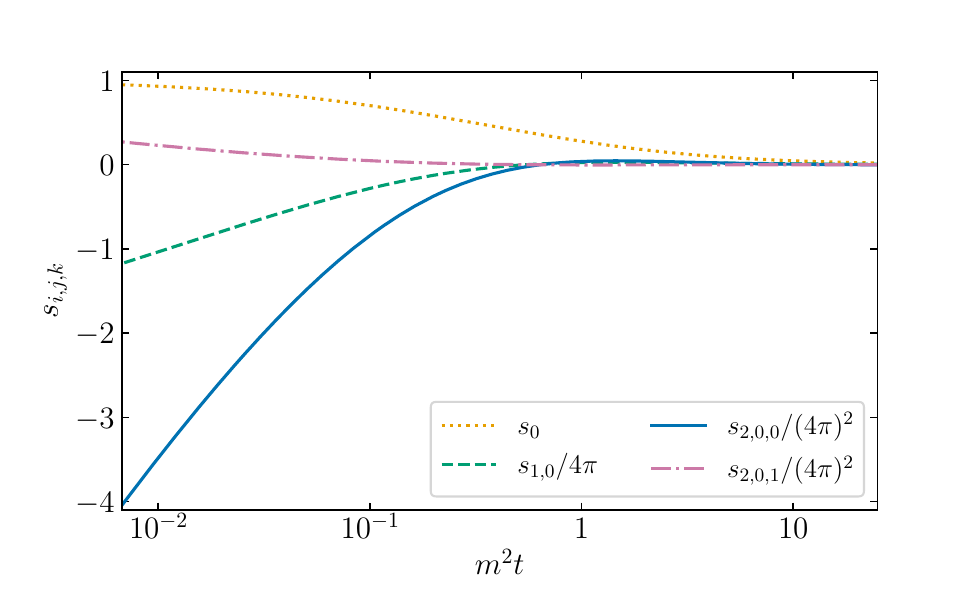} 
    \caption{$\lmut$-independent components of $\mathring{S}_f(t)$ in the form
      of \cref{eq:Resu:jael,eq:Resu:epic}.
      \label{fig:SCOMP3L}}
\end{figure}


\cref{fig:SCOMP3L} shows the non-logarithmic terms $s_{0,0}$, $s_{1,0}$,
$s_{2,0,0}$, and $s_{2,0,1}$. For small $m^2t$, the change in the components
is generally very steep, where the overall process is expected to be linear in
the mass value. It then flattens out for large mass where it is expected to
vary as $1/m^2t$.  A comparison of the components of $s_{2,i,0}$ and
$s_{2,i,1}$ suggests that the contributions from the secondary massless quarks
appearing in loops is subdominant to the massive quark.

Switching to a different renormalization scheme for the quark mass can be
achieved as follows. Assume that the relation between the on-shell mass $m$
and the mass in the new scheme $\hat{m}$ is given by
\begin{equation}\label{eq:Resu:brix}
  \begin{aligned}
    m = \hat{m}\left(1 + \api\hat{\delta}_m^{(1)} + \api^2\hat{\delta}_m^{(2)} +
    \ldots\right)\,.
  \end{aligned}
\end{equation}
For the \msbar\ scheme, $\hat{m} = \bar{m}(\mu)$ and $\hat{\delta}_m^{(n)} =
\delta_m^{(n)}$, see \cref{eq:Reno:floc}, for example.  Then
$\mathring{S}_f(t)$ is given by the same formulas as \cref{eq:Resu:jael} with
$m_f$ replaced by $\hat{m}_f$, and $s_n$ replaced by $\hat{s}_n$, where
\begin{equation}\label{eq:Resu:enos}
  \begin{aligned}
    \hat{s}_0 &= s_0\,,\\ \hat{s}_1 &= s_1 + \delta_m^{(1)}\left(s_0 +
    2\hat{z}\,s'_0\right)\,,\\ \hat{s}_2 &= s_2 + \delta_m^{(1)}\left(s_1 +
    2\hat{z}\,s'_1\right) + \delta_m^{(2)}\left(s_0 + 2\hat{z}\,s'_0\right) +
    \hat{z}\,(\delta_m^{(1)})^2 \left(3s'_0 + 2\hat{z}\,s''_0\right)\,,\\
    \hat{z} &= \hat{m}^2t\,,
  \end{aligned}
\end{equation}
and
\begin{equation}\label{eq:Resu:circ}
  \begin{aligned}
    s'_{n,k}(\hat{z}) &= \dderiv{}{}{\hat{z}}s_{n,k}(\hat{z})\,.
  \end{aligned}
\end{equation}


\begin{figure}[!ht]
    \centering
    \includegraphics[width=0.9\textwidth]{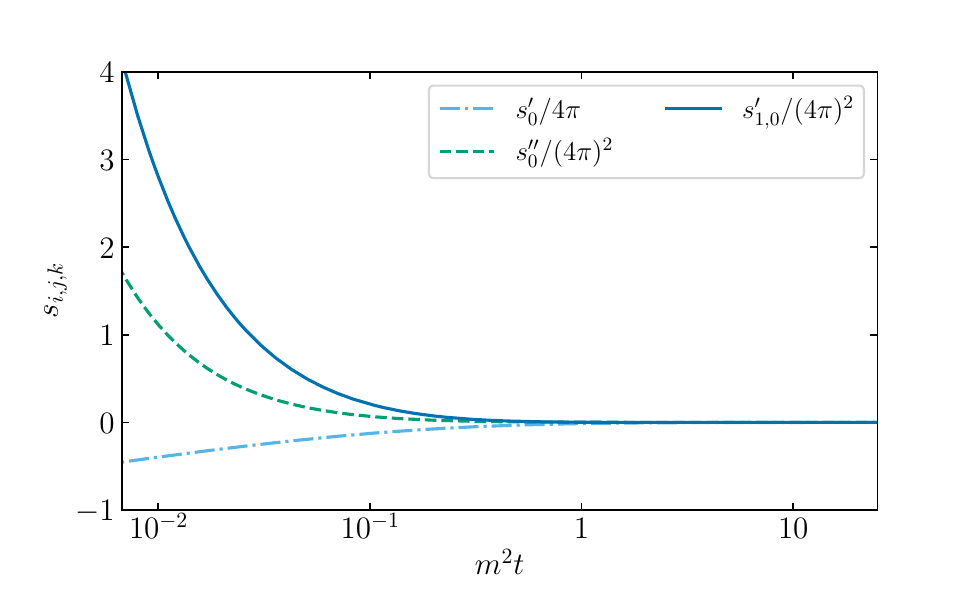} 
    \caption{Derivatives of the $\lmut$-independent components of
      $\mathring{S}_f(t)$. \label{fig:SCOMPDER3L}}
\end{figure}


Thus, in order to convert between different mass schemes, aside from the
non-logarithmic terms, we also need the first and second derivative of the
\lo\ term, which is easily obtained from \cref{eq:Resu:fief}, as well as the
first derivative of the \nlo\ coefficient, which we provide in terms of a
\one-dimensional grid along with the non-logarithmic coefficients; it is also
plotted in \cref{fig:SCOMPDER3L}.


\subsection{The quark kinetic density}\label{sec:rres}


\begin{figure}[!ht]
    \centering
    \includegraphics[width=0.9\textwidth]{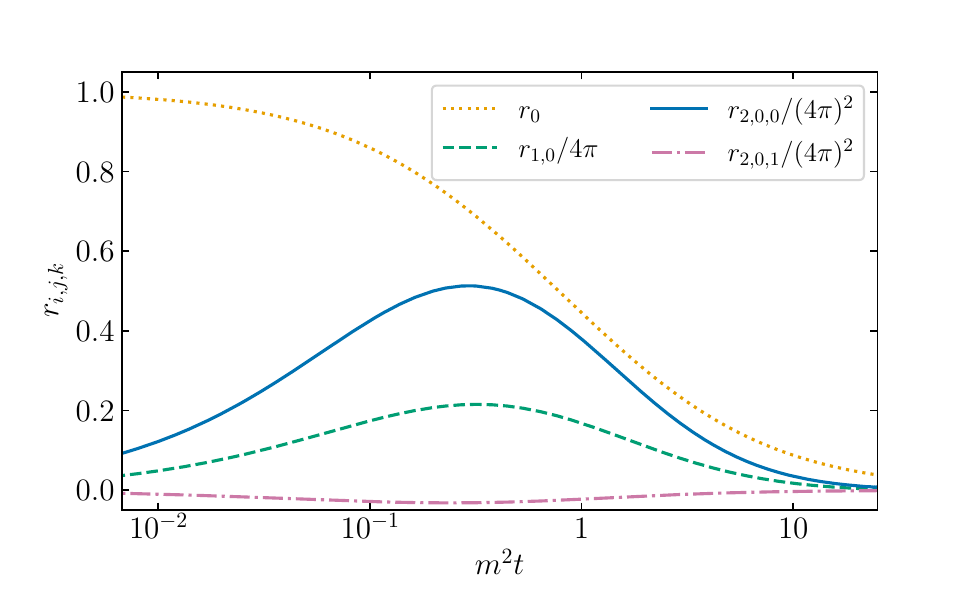}
    \caption{$\lmut$-independent components of $\mathring{R}_f(t)$ in the form
      \cref{eq:RtEXP}.}\label{fig:RCOMP3L}
\end{figure}


\begin{figure}[!ht]
    \centering
    \includegraphics[width=0.9\textwidth]{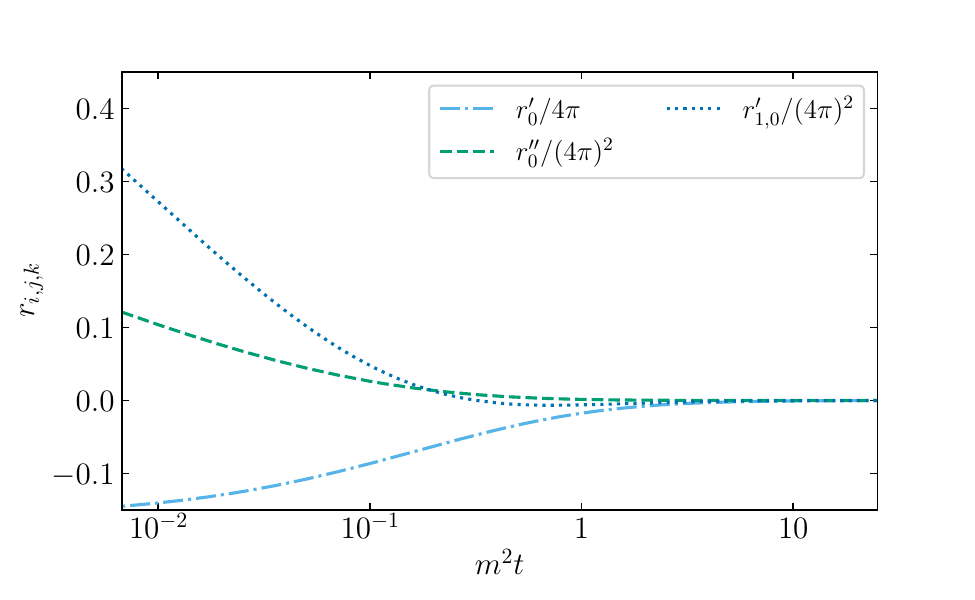}
    \caption{$m^2t$ derivatives of $\lmut$-independent components of
      $\mathring{R}_f(t)$.}
    \label{fig:RCOMPDER3L}
\end{figure}


In analogy to \cref{eq:Resu:jael}, we write the \vev\ of the quark kinetic
density in the ringed scheme up to order $N$ as
\begin{equation}\label{eq:RtEXP}
  \begin{aligned}
    \mathring{R}_f(t) &= -\frac{2\nc}{(4\pi t)^2} \sum_{n=0}^N\api^n
    r_n(m_f,t,\mu)\,, \qquad
    r_n(m,t,\mu) = \sum_{k=0}^n
    r_{n,k}(m^2 t)\lmut^k\,.
  \end{aligned}
\end{equation}
Again, we further write
\begin{equation}\label{eq:Resu:ifni}
  \begin{aligned}
    r_{2,k} &= r_{2,k,0} + \nl\,r_{2,k,1}\,.
  \end{aligned}
\end{equation}
As for $\mathring{S}_f$, a logarithmic term only appears at \nnlo, and it can
be expressed by the non-logarithmic \nlo\ coefficient:
\begin{equation}\label{eq:Resu:hookr}
  \begin{aligned}
    r_{2,1} = \beta_0\, r_{1,0}\,.
  \end{aligned}
\end{equation}
The \lo\ coefficient is given by
\begin{equation}\label{eq:Resu:alex}
  \begin{aligned}
    r_{0,0}(z) &= 1-z+4z^2\,e^{2z}\Gamma(0,2z)\,.
  \end{aligned}
\end{equation}
The dimensionless coefficients $r_{0,0}$, $r_{1,0}$, $r_{2,0,0}$ and
$r_{2,0,1}$ are plotted in \cref{fig:RCOMP3L}.

The conversion to another mass scheme as defined in \cref{eq:Resu:brix} in
this case is obtained by replacing $m$ by $\hat{m}$ and $r_{n}$ by
$\hat{r}_n$, where
\begin{equation}\label{eq:Resu:jhwh}
  \begin{aligned}
    \hat{r}_0 &= r_0\,,\\ \hat{r}_1 &= r_1
    + 2\,\delta_m^{(1)}\hat{z}\,r'_0\,,\\
    \hat{r}_2 &= r_2 + 2\,\delta_m^{(1)}
    \hat{z}\,r'_1 +
    \hat{z}\,\left(
    (\delta_m^{(1)})^2 + 
    2\,\delta_m^{(2)}\right)r'_0 +
    2(\delta_m^{(1)})^2\hat{z}^2r''_0\,.
      \end{aligned}
\end{equation}
The derivatives $r_0'$, $r_1'=r_{1,0}'$, and $r_{0}''$ are shown in
\cref{fig:RCOMPDER3L}.


\subsection{The action density}\label{sec:eres}


\begin{figure}[!ht]
  \centering
    \includegraphics[width=0.9\textwidth]{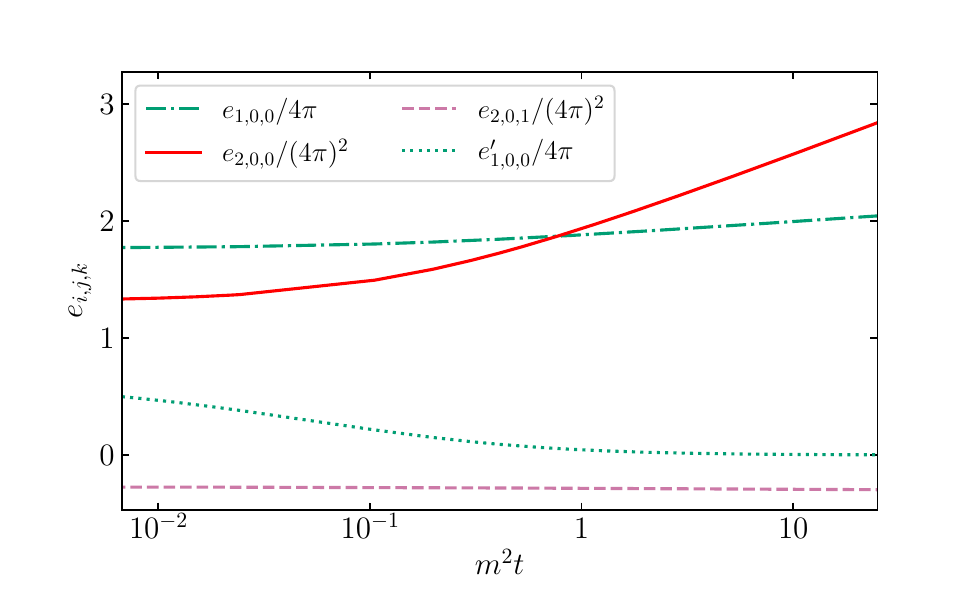}
    \caption{Components of $ E(t) $ in the form \cref{eq:EtEXP} and its 
    $m^2t$ derivatives of $\lmut$-independent components.
      \label{fig:ECOMP3L}}
\end{figure}


The \vev\ of the action density $ E(t) $ is expanded as
\begin{equation}\label{eq:EtEXP}
  \begin{aligned}
    E(t) &= \frac{3\alpha_s}{4\pi t^2}\frac{\na}{8}
    \sum_{n=0}^N \api^n e_{n}(m,t,\mu)\,,\qquad
    e_n(m,t,\mu) = \sum_{k=0}^n e_{n,k}(m^2t)\lmut^k\,,
  \end{aligned}
\end{equation}
where $\na=\nc^2-1=8$ is the dimension of the adjoint representation of SU(3).  In this case, already the \nlo\ coefficients depend on
the number of light fermions, so we write
\begin{equation}\label{eq:Resu:kana}
  \begin{aligned}
    e_{1,k} &= e_{1,k,0} + \nl\, e_{1,k,1}\,,\qquad
    e_{2,k} &= e_{2,k,0} + \nl\, e_{2,k,1} + \nl^2\, e_{2,k,2}\,.
  \end{aligned}
\end{equation}
Since $E(t)$ starts at order $\alpha_s$, logarithmic terms appear already at
\nlo{}. They can be expressed in terms of the lower-order non-logarithmic
terms as
\begin{equation}\label{eq:Resu:iron}
  \begin{aligned}
    e_{1,1} &= \beta_0 e_{0,0}\,,\qquad
    e_{2,1} = \beta_1 e_{0,0} + 2\beta_0 e_{1,0}\,,\qquad
    e_{2,2} = 2\beta^2_0 e_{0,0}\,,\\
    e_{n,k} &= 0\quad\text{otherwise, for}\quad n\leq 2\ \wedge\ k\geq 1\,,
  \end{aligned}
\end{equation}
where $\beta_0$ and $\beta_1$ are given in \cref{eq:Resu:byng}.  The
\lo\ result is simply given by
\begin{equation}\label{eq:Resu:baya}
  \begin{aligned}
    e_{0,0}(z) &= 1\,.
  \end{aligned}
\end{equation}
The coefficient of the highest power of $\nl$ at each order in $\alpha_s$ is a
constant in $m^2t$.  These values can be directly compared to the
corresponding massless coefficients in
\citeres{Luscher:2010iy,Harlander:2016vzb,Artz:2019bpr}
\begin{equation}\label{eq:Resu:brno}
  \begin{aligned}
    e_{1,0,1} &= -\frac{4}{9}\,\qquad\text{and}\qquad
    e_{2,0,2} = -\frac{20}{81} + \frac{2\pi^2}{27}\,.
  \end{aligned}
\end{equation}
The other independent components, $e_{1,0,0}$, $e_{2,0,0}$ and $e_{2,0,1}$ are plotted in
\cref{fig:ECOMP3L}.

Conversion from the on-shell to the \msbar\ mass is particularly simple in
this case:
\begin{equation}\label{eq:Resu:jhwh1}
  \begin{aligned}
    \hat{e}_0 &= e_0\,,\qquad \hat{e}_1 = e_1\,,\qquad
    \hat{e}_2 = e_2 + 2\,\delta_m^{(1)} \hat{z}\,e'_1\,.
  \end{aligned}
\end{equation}
The only mass-dependent component of $e'_1$, $e'_{1,0,0}$ is also plotted in
\cref{fig:ECOMP3L}.



\section{Phenomenological results}\label{sec:pheno}

While in \cref{sec:sres,sec:rres,sec:eres}, we provided the individual
components of the \vevs{} as functions of $m^2t$, we will now combine them in
order to obtain numerical values for the complete \vevs{} up to \nnlo\ \qcd.
In particular, we will estimate their theoretical accuracy by varying the
renormalization scale.

As pointed out above, we will consider the cases of \qcd\ with $\nf=4$ and
$\nf=5$ quark flavors, where $\nf-1$ quarks are taken to be massless, so that
the mass effects only arise from the heaviest quark. Thus, for $\nf=4$
($\nf=5$), we will only consider $S_f(t)$ and $R_f(t)$ for $f=c$ ($f=b$),
cf.\ \cref{eq:Flow:bank}. We will briefly study the effect of a second massive
quark at the end of this section.

Throughout this paper, our input value for the strong coupling is
\begin{equation}\label{eq:Resu:gawk}
  \begin{aligned}
    \alpha_{s}(M_Z)=0.118\quad \text{with}\quad M_Z=91.188\,\text{GeV}\,.
  \end{aligned}
\end{equation}
For the numerical results, we renormalize the quark mass in the
\msbar\ scheme, using
\begin{equation}\label{eq:Resu:flon}
  \begin{aligned}
    \bar{m}_c(\bar{m}_c)=1.273\,\text{GeV}~~~\text{and}~~~
    \bar{m}_b(\bar{m}_b)=4.183\,\text{GeV}
  \end{aligned}
\end{equation}
as input values~\cite{ParticleDataGroup:2024cfk}.  For each of these masses,
we follow \citere{Takaura:2025pao} and define a central renormalization scale
\begin{eqnarray}\label{eq:res:muint}
\mu_\text{int} = \sqrt{\mu_t^2 + \bar{m}^2(\bar{m})}\,,\quad\mathrm{where}\quad
\mu_t=\frac{\exp(-\frac{\EulerGamma}{2})}{\sqrt{2t}}\,,
\end{eqnarray}
which aims at avoiding large logarithms emerging due to the difference between
the quark mass and the flow time. For the case of the gluon condensate $E(t)$,
the central scale $\mu_t$ is used since the contributions from the quark mass
first appear at the two loop level, as such the logarithms from the mass scale
are suppressed.


\begin{figure}[!ht]
  \begin{center}
    \begin{tabular}{cc}
    \includegraphics[width=0.47\textwidth]{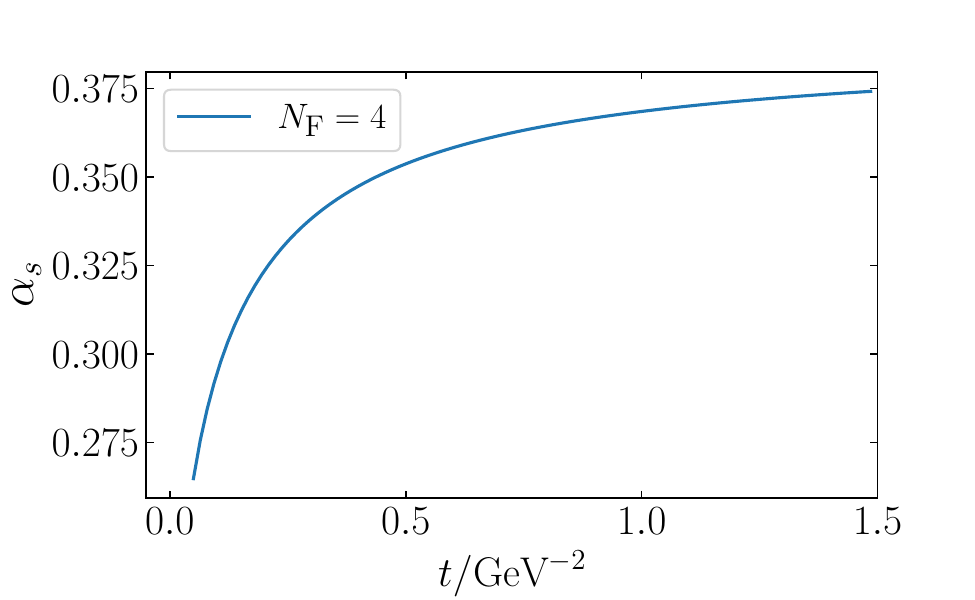} &
    \includegraphics[width=0.47\textwidth]{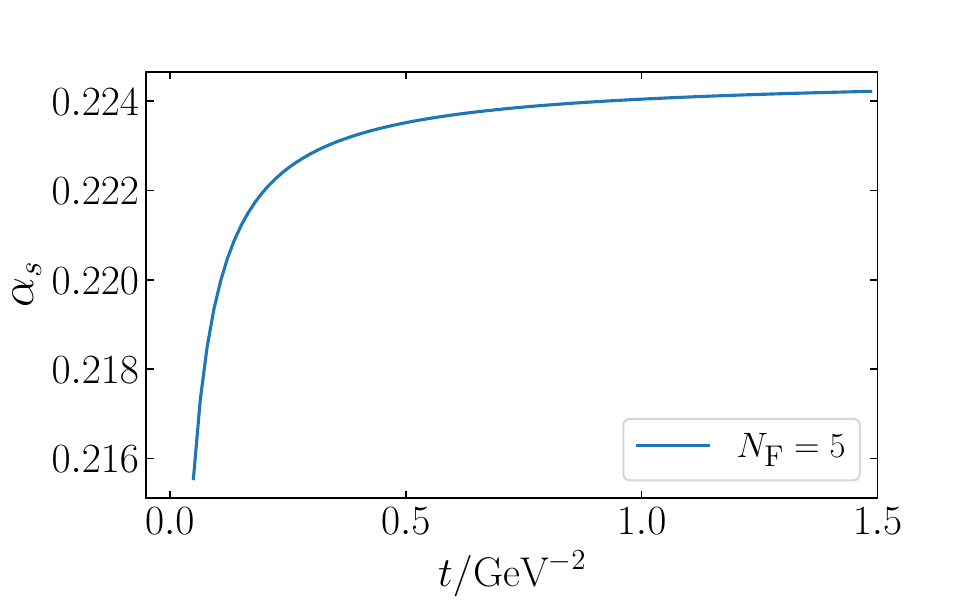}\\
    (a) & (b)\\
    \includegraphics[width=0.47\textwidth]{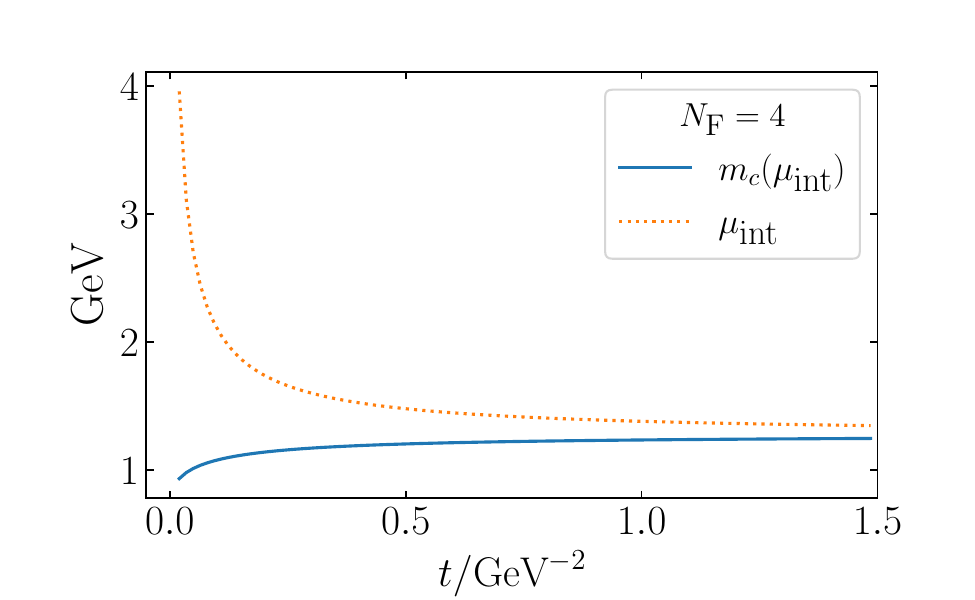} &
    \includegraphics[width=0.47\textwidth]{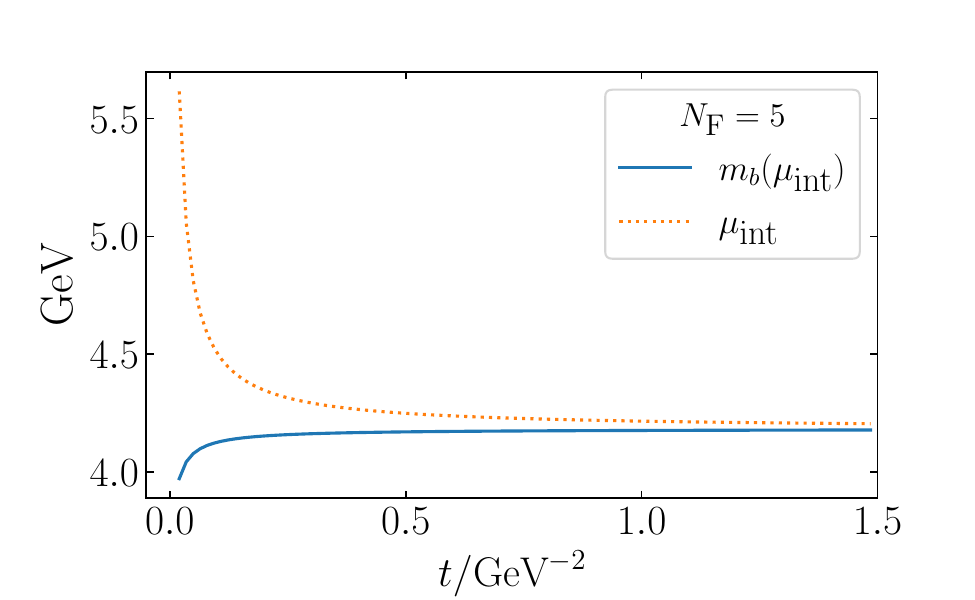} \\
    (c) & (d)
    \end{tabular}
    \caption{Upper row (a/b): $\alpha(\mu_{\text{int}})$.  Lower row (c/d):
      $m_f(\mu_\text{int})$ (solid)
      and $\mu_\text{int}$ (dotted). Left column (a/c): $\nf=4$;
      right column (b/d): $\nf=5$.
    \label{fig:running}}
  \end{center}
\end{figure}


For each quantity, $\alpha_s$ is evolved from the reference scale $M_Z$ to the
central scale $\mu_\text{int}$ at the \five-loop level. When working in the
($\nf=4$)-flavor theory, we use the \four-loop decoupling relations for
$\alpha_s$ at $\mu=\bar{m}_b(\bar{m}_b)$.  Similarly, we evolve the quark
masses from $\bar{m}(\bar{m})$ to $\bar{m}(\mu_\text{int})$ at \five-loop
level. To provide an understanding of the scales we are considering, in
\cref{fig:running} we plot the running coupling constant (upper row) and the
running charm and bottom mass (lower row) as functions of the flow time $t$,
both for $\nf=4$ (left) and $\nf=5$ (right) as a function of $t$.  These plots
demonstrate the regularization of non-perturbative behavior which occurs in
the massless limit towards large $t$. Due to the occurrence of the quark mass
in $\mu_\text{int}$, the coupling and the quark mass tend towards constants in
this limit.


\subsection{Size of the mass effects}\label{sec:size}


\begin{figure}[!ht]
  \begin{center}
    \begin{tabular}{cc}
    \includegraphics[width=0.47\textwidth]{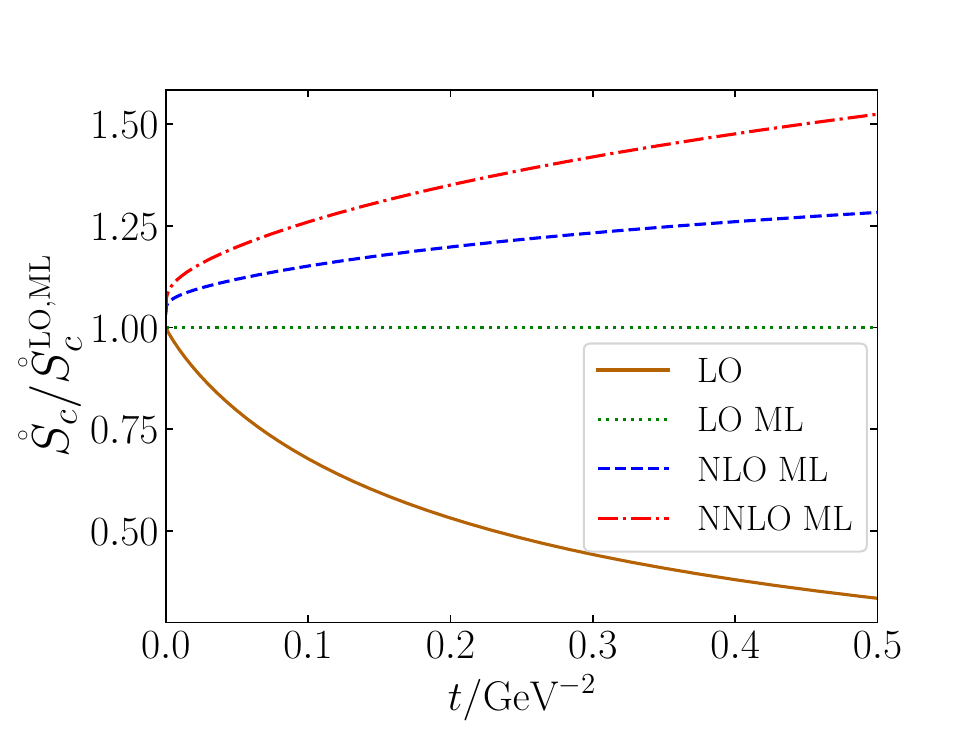} &
    \includegraphics[width=0.47\textwidth]{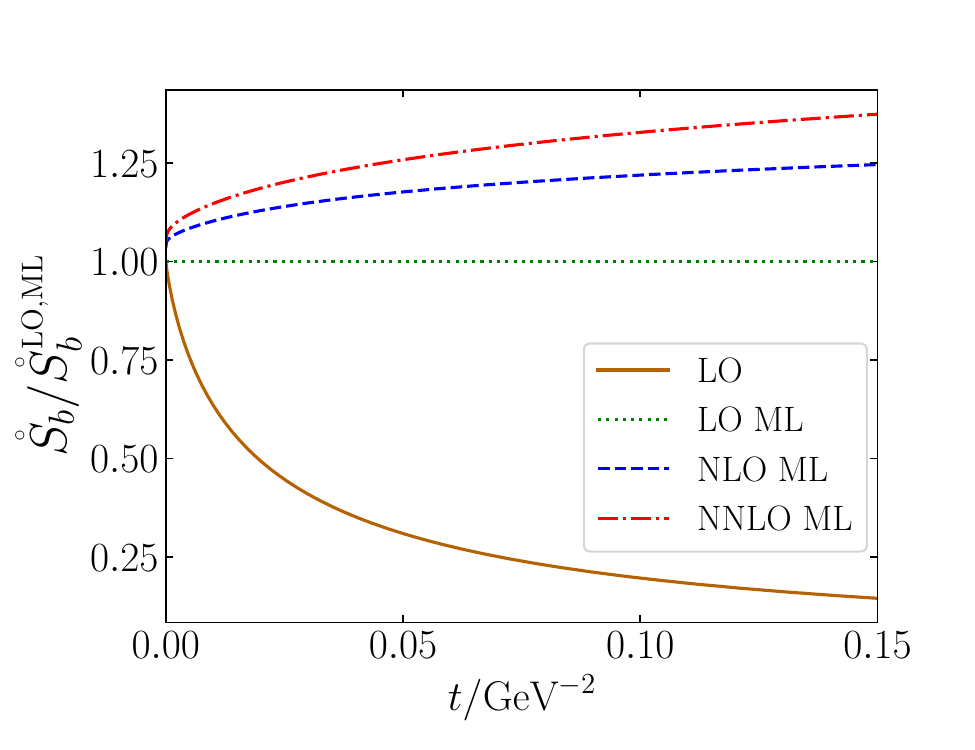}\\
    (a) & (b)\\
    \includegraphics[width=0.47\textwidth]{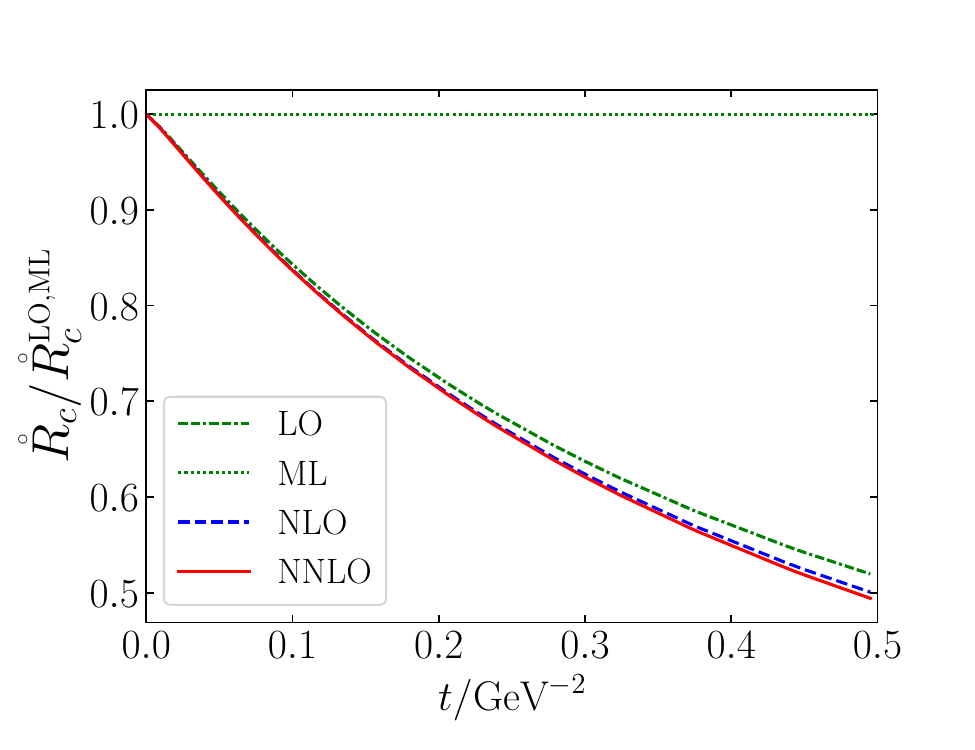} &
    \includegraphics[width=0.47\textwidth]{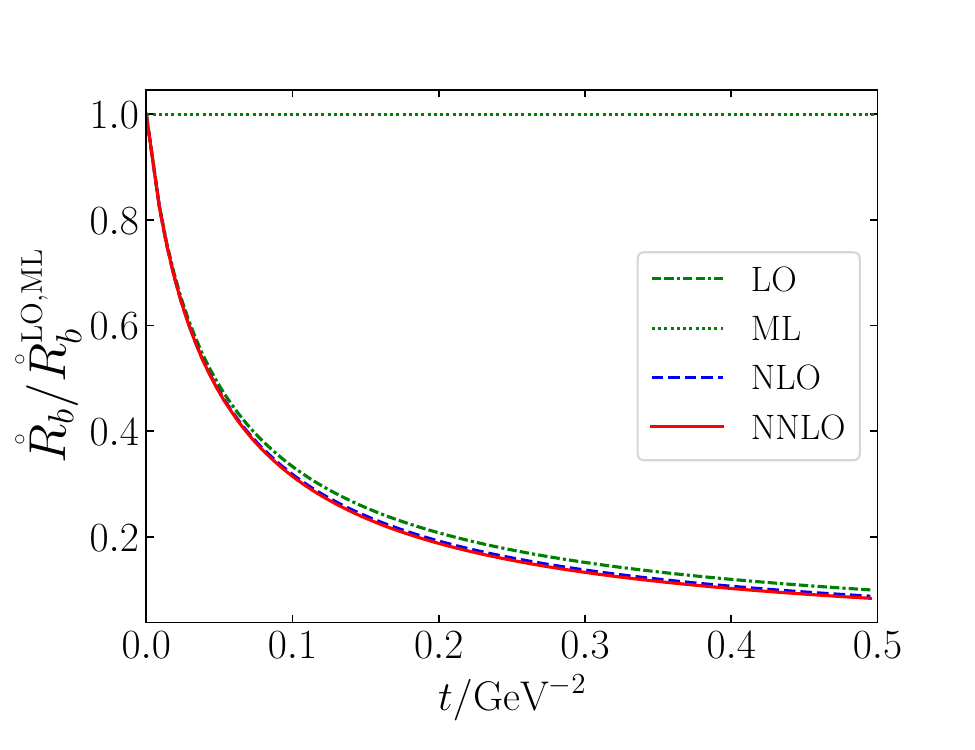}\\
    (c) & (d)
    \end{tabular}
    \caption{Upper row (a/b):
      $\mathring{S}_f/\mathring{S}^\text{\lo,\abbrev{ML}}_f$, where
      $\mathring{S}_f$ is taken in the ``massless'' limit (\abbrev{ML},
      cf.\,\cref{eq:defsfml}) at \lo, \nlo, and \nnlo\ perturbation theory
      (dotted, dashed, dash-dotted), and the \lo\ result including effects
      from non-zero $m_f$ (solid); left (a/c): $f=c$, right (b/d): $f=b$.
      Lower row (c/d): analogous results for
      $\mathring{R}_f/\mathring{R}^\text{\lo,\abbrev{ML}}_f$; note that in
      this case, higher perturbative orders in the massless case are zero by
      definition of the ringed scheme; instead, the \nlo\ and \nnlo\ curves
      are included here, evaluated at the central scale $\mu_\text{int}$.
      \label{fig:MLCOMP}}
  \end{center}
\end{figure}


One of the primary purposes of this work is to evaluate the importance of the
quark-mass effects in these gradient-flow quantities. The first question we
will ask therefore is how the \vevs{} compare to their massless
equivalents. For the case of $E(t)$ and $R_f(t)$ we will simply use the
massless values of the series. However, since $S_f(t)\propto m_f$, the na\"{i}ve
massless equivalent vanishes. Instead, we will consider
\begin{equation}
  \begin{aligned}
  \mathring{S}^\text{\abbrev{ML}}_f(t) = m_f\left[ \left.\frac{d
      \mathring{S}_f(t)}{dm_f}\right|_{m_f=0} \right].
  \label{eq:defsfml}
  \end{aligned}
\end{equation}
This quantity is used in \citere{Artz:2019bpr} to define the gradient-flow
mass; it is equivalent to the leading term in the small-mass expansion.
\cref{fig:MLCOMP} compares successively higher perturbative orders of this
massless limit to the leading-order result which, however, includes the full
mass dependence. It shows that the quark-mass effects are considerably larger
than the radiative corrections already at rather small values of $t$.  In
fact, as noted in \citere{Takaura:2025pao}, even including
higher terms in an expansion around the small-mass limit has only a very
limited range of applicability.

The analogous comparison for $\mathring{R}_f$ is shown in the lower row of
\cref{fig:MLCOMP}.  By definition of the ringed scheme, there are no
higher-order corrections in the massless limit in this case, see
\cref{eq:Reno:hern}. The relative size of the \lo\ mass effects is comparable
to those in $\mathring{S}_f$. \cref{fig:MLCOMP}~(c) and~(d) also include
higher orders for the full massive results. Compared to the \lo\ mass effects
as well as to the radiative corrections in the massless case, we find these
corrections to be very moderate. They will be studied in more detail below for
all \vevs\ under consideration.


\begin{figure}[!ht]
  \begin{center}
    \begin{tabular}{cc}
    \includegraphics[width=0.47\textwidth]{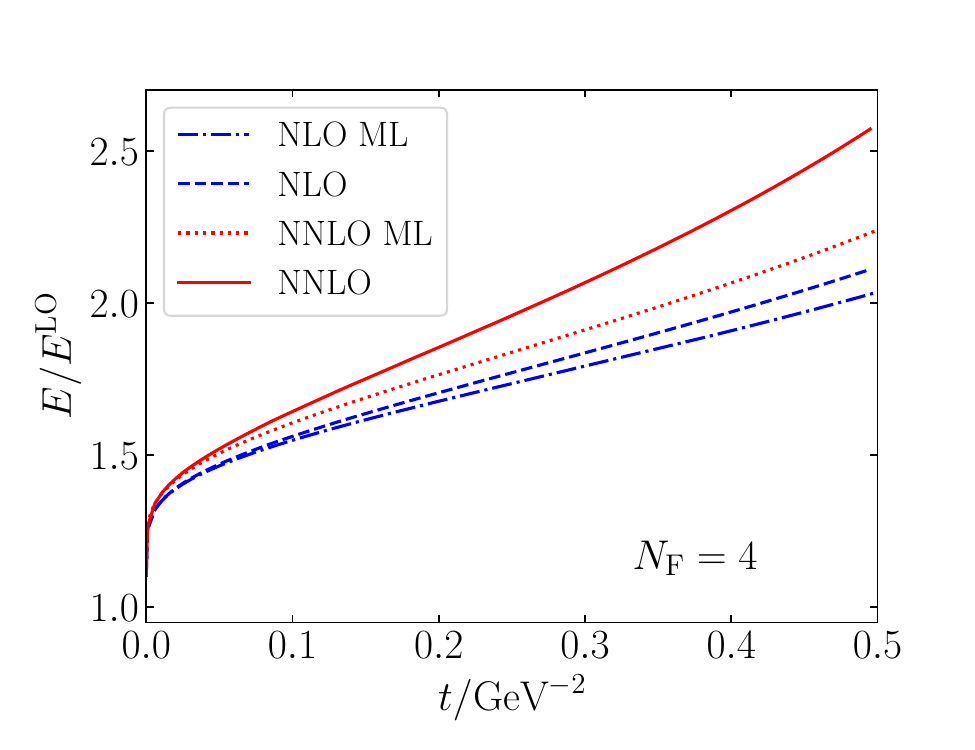} &
    \includegraphics[width=0.47\textwidth]{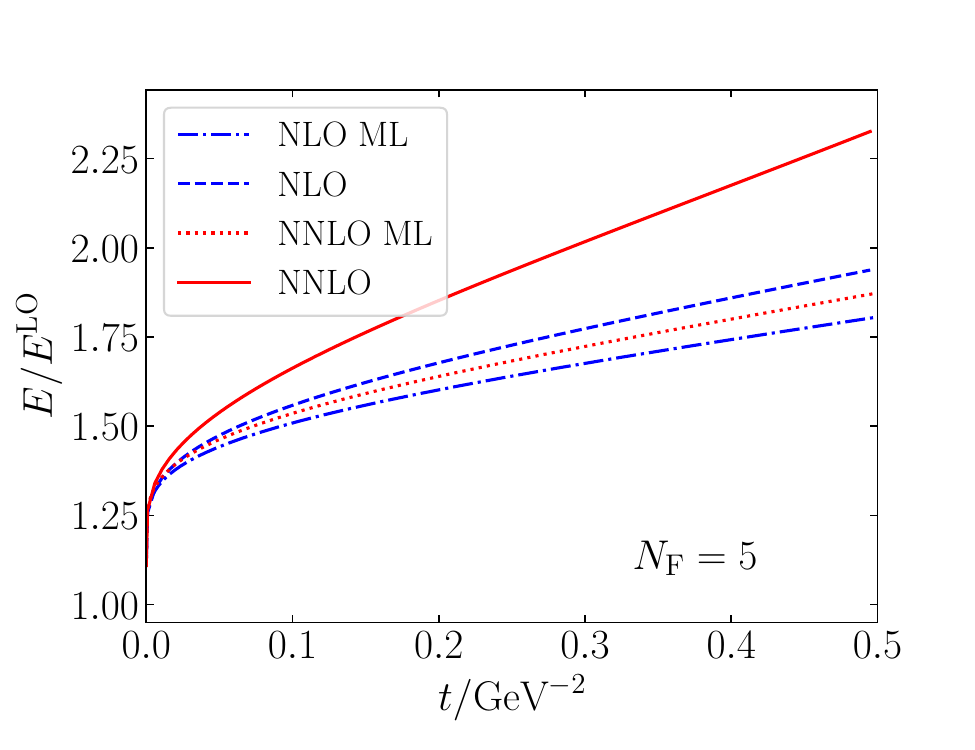}\\
    (a) & (b)
    \end{tabular}
    \caption{$E/E^\text{\lo}$ for $\nf=4$ (left) and
      $\nf=5$ (right) at \nlo{} and \nnlo{} with either one or zero massive
      quarks. }\label{fig:EMSMLCOMP}
  \end{center}
\end{figure}


Finally, \cref{fig:EMSMLCOMP} compares $E(t)$ including the full quark-mass
dependence to the case of purely massless quarks. In this case, there are no
mass effects in the \lo\ result as is obvious from the contributing Feynman
diagrams, see \cref{fig:GenFeynmanDiagrams}. While the mass effects at
\nlo\ are quite moderate, amounting to less than 4\% below
$t=0.5/\text{GeV}^2$ in the case of a massive charm quark (and $\nf=4$,
\cref{fig:EMSMLCOMP}\,(a)), they reach up to 15\% at \nnlo. In the
$(\nf=5)$-case with a massive bottom quark, these numbers are 6\% and 20\%,
respectively, see \cref{fig:EMSMLCOMP}\,(b).


\subsection{Radiative corrections}\label{sec:radcor}

Let us now study the radiative corrections to the \vevs{} when the full mass
effects are included. In particular, we want to address the residual
perturbative uncertainty estimated by varying the renormalization scale.  For
this variation, we evolve the mass and the coupling according to the
perturbative order of the \vevs{}. As in both $\mathring{S}_f(t)$ and
$\mathring{R}_f(t)$ the mass appears at \lo{}, we evolve $m(\mu_\text{int})$
to $m(\mu)$ at $N$-loop level when considering these quantities at $N$-loop
order ($N=1,2,3$). On the other hand, $\alpha_s$ first appears at \nlo, which
is why we evolve $\alpha_s(\mu_\text{int})$ to $\alpha_s(\mu)$ only at
$(N-1)$-loop level. For $E(t)$, on the other hand, the situation is reversed:
the coupling appears at \lo{} and the mass only at \nlo{}, so we evolve
$\alpha_s$ at $N$-loop level and the mass at $(N-1)$-loop level in this case.
All running and decoupling calculations are done with the help of
\verb|RunDec|~\cite{Chetyrkin:2000yt,Schmidt:2012az,Herren:2017osy}.


\begin{figure}[!ht]
  \begin{center}
    \begin{tabular}{cc}
    \includegraphics[width=0.47\textwidth]{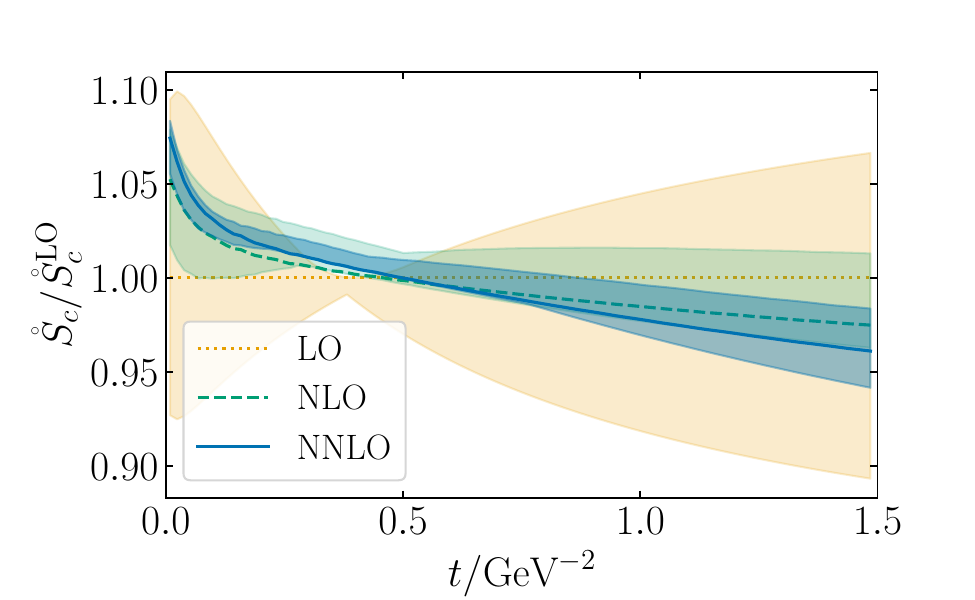} &
    \includegraphics[width=0.47\textwidth]{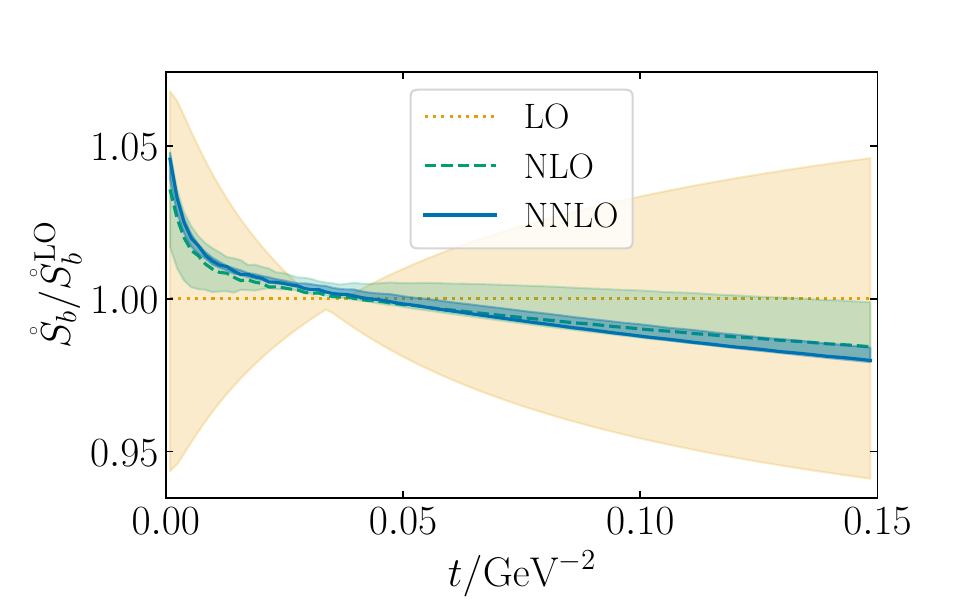}\\
    (a) & (b)\\
    \includegraphics[width=0.47\textwidth]{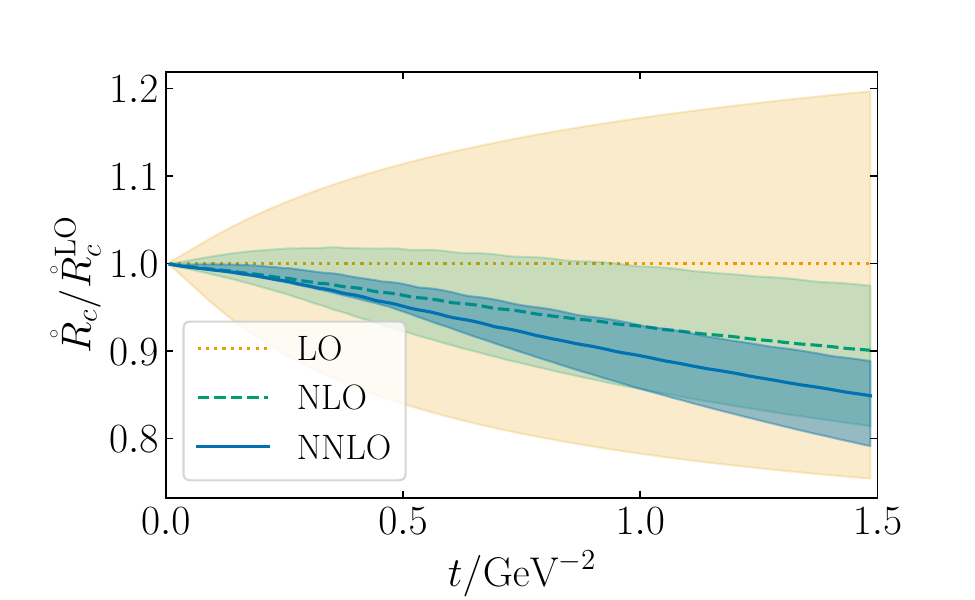} &
    \includegraphics[width=0.47\textwidth]{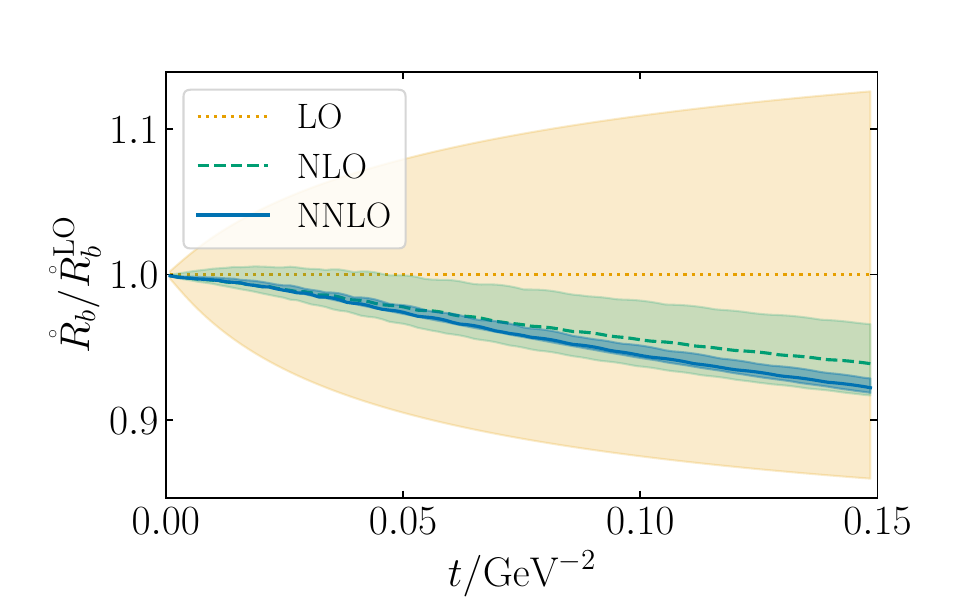}\\
    (c) & (d)
    \end{tabular}
    \caption{Upper row (a/b): $\mathring{S}_f/\mathring{S}_f^\text{\lo}$ at
      \nlo{} and \nnlo{}, including full quark-mass effects, with an envelope
      error formed by scale variation, see
      \cref{eq:Resu:berm,eq:Resu:ecsc,eq:Resu:jail}. Lower row (c/d): The
      analogous plots for $\mathring{R}_f/\mathring{R}_f^\text{\lo}$. Left
      column (a/c): $f=c$, i.e.\ $\nf=4$ and massive charm quark; right column
      (b/d): $f=b$, i.e.\ $\nf=5$ and massive bottom quark.
    \label{fig:SMSCOMPHalf}}
  \end{center}
\end{figure}


For each value in $t$, the scale $\mu$ is varied in the range
\begin{equation}\label{eq:Resu:berm}
  \begin{aligned}
    \mu\in\Delta_\mu = [\mu_\text{int}/2,2\,\mu_\text{int}]\,.
  \end{aligned}
\end{equation}
For a measurable quantity $\rho$, the limits of the scale envelope are the
maximum and minimum values in this range, i.e.\ for each value of $t$, it is
\begin{equation}\label{eq:Resu:ecsc}
  \begin{aligned}
    \rho&={\rho_0}^{+\delta\rho_+}_{-\delta\rho_-}
  \end{aligned}
\end{equation}
where 
\begin{equation}\label{eq:Resu:jail}
  \begin{aligned}
    \rho_0 &= \rho|_{\mu=\mu_\text{int}}\,,\qquad
    \delta\rho_+ = \max_{\mu\in\Delta_\mu}\rho-\rho_0\,,\qquad
    \delta\rho_- &= \rho_0-\min_{\mu\in\Delta_\mu}\rho\,,
  \end{aligned}
\end{equation}
The upper row of \cref{fig:SMSCOMPHalf} displays $\mathring{S}_f(t)$,
evaluated at various loop orders and normalized to the \lo\ expression taken
at the central scale $\mu_\text{int}$. The shaded regions indicate the
perturbative uncertainties as given by \cref{eq:Resu:ecsc}. In the left plot,
\cref{fig:SMSCOMPHalf}\,(a), we consider $\mathring{S}_c(t)$ in the
\four-flavor theory ($\nf=4$), keeping only the charm-quark mass different
from zero. The right plot shows $\mathring{S}_b(t)$ for $\nf=5$, and only the
bottom quark is considered massive. The expected reduction of the theoretical
uncertainty when going to higher orders is particularly apparent in the case
of $\nf=5$. Note that we use different ranges of~$t$ because our choice of the
central scale $\mu_\text{int}$ depends on the mass of the massive fermion.

In general, the \nnlo\ corrections lead to a considerable reduction of the
theoretical uncertainty in both cases. While for the charm quark case, the
uncertainty bands of the successive orders overlap over the whole range in $t$
shown in the plots, the \nlo\ and \nnlo\ bands become slightly incompatible in
the case of the bottom quark for $t\gtrsim 0.6\,\mathrm{GeV}^{-2}$, albeit
still to an acceptable extent.
The analogous plots for $\mathring{R}_f/\mathring{R}_f^\text{\lo}$ are shown
in the lower row of \cref{fig:SMSCOMPHalf}.  For $m_f \to 0$ (and thus $t\to
0$), the ratio tends to~1 in this case due to the definition of the ringed
scheme.


\begin{figure}[!ht]
  \centering
     \begin{center}
    \begin{tabular}{cc}
      \includegraphics[width=0.47\textwidth]{%
        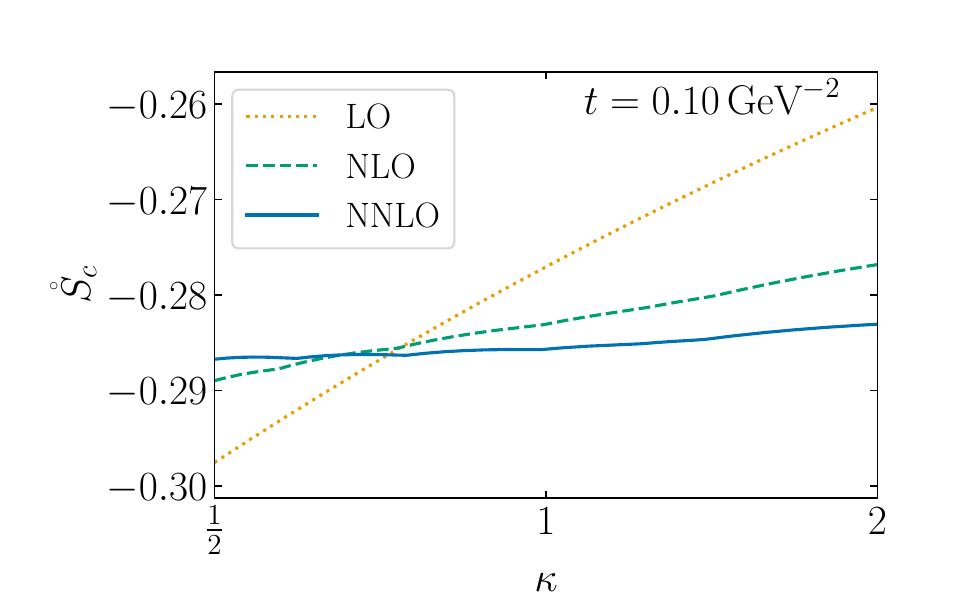} &
      \includegraphics[width=0.47\textwidth]{%
        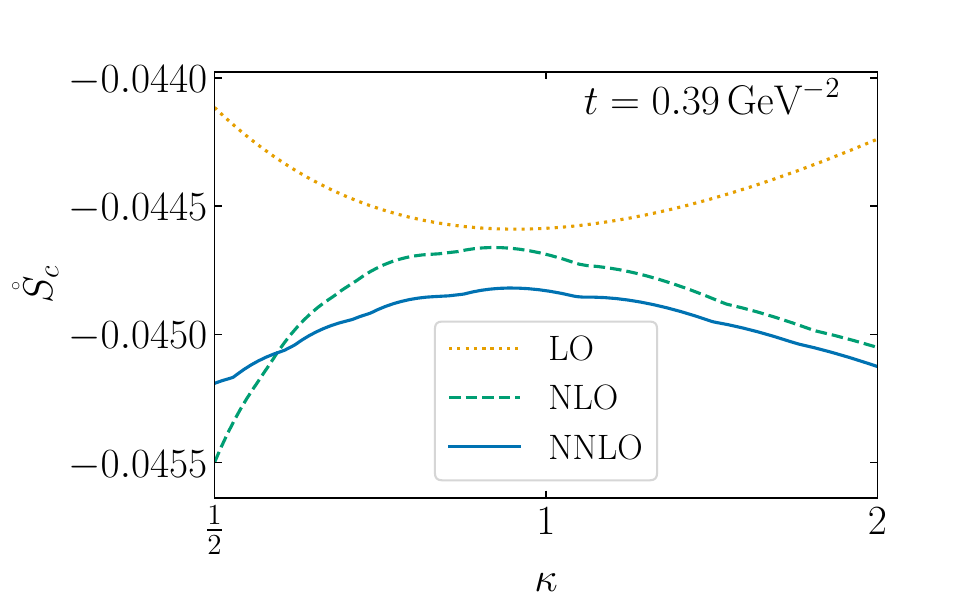}\\
      (a)& (b) \\
      \includegraphics[width=0.47\textwidth]{%
        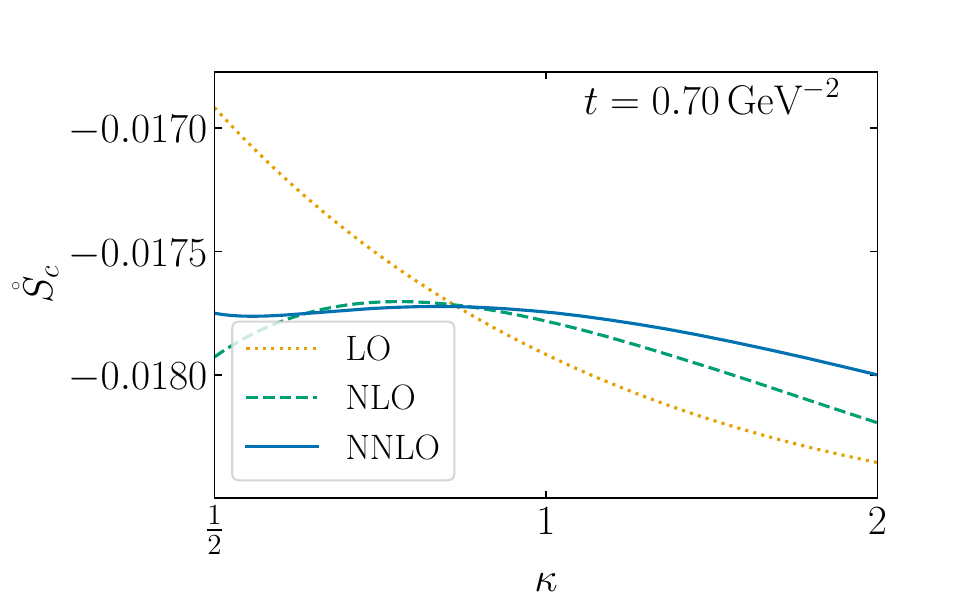} &
      \includegraphics[width=0.47\textwidth]{%
        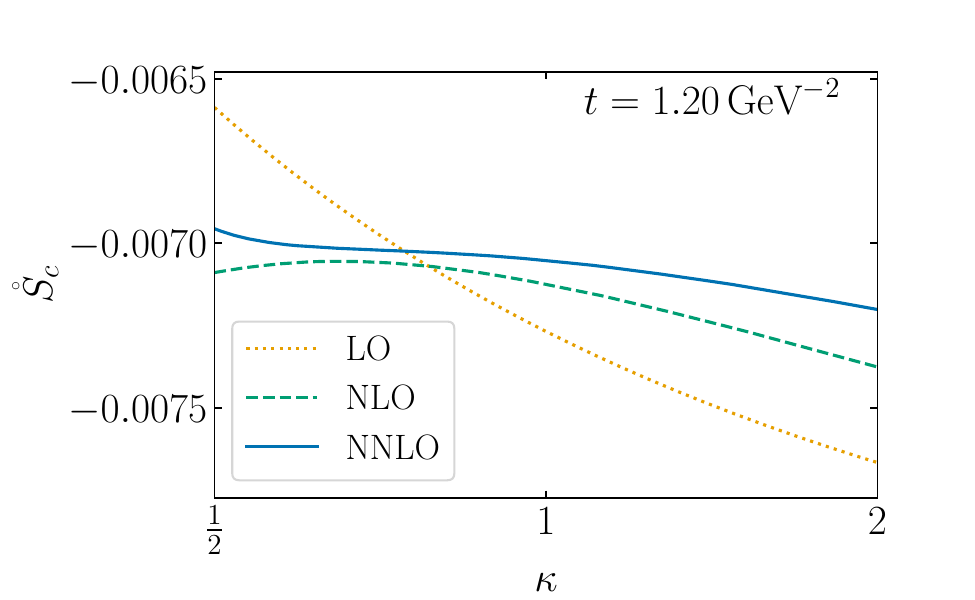} \\
      (c) & (d)
    \end{tabular}
  \end{center}
    \caption{$\mathring{S}_c$ plotted across a range of
      $\kappa=\mu/\mu_{\text{int}}$ for various values of the flow time $t$.
      \label{fig:raMuComp}}
  \end{figure}


\begin{figure}[!ht]
  \begin{center}
    \begin{tabular}{cc}
      \includegraphics[width=0.47\textwidth]{%
        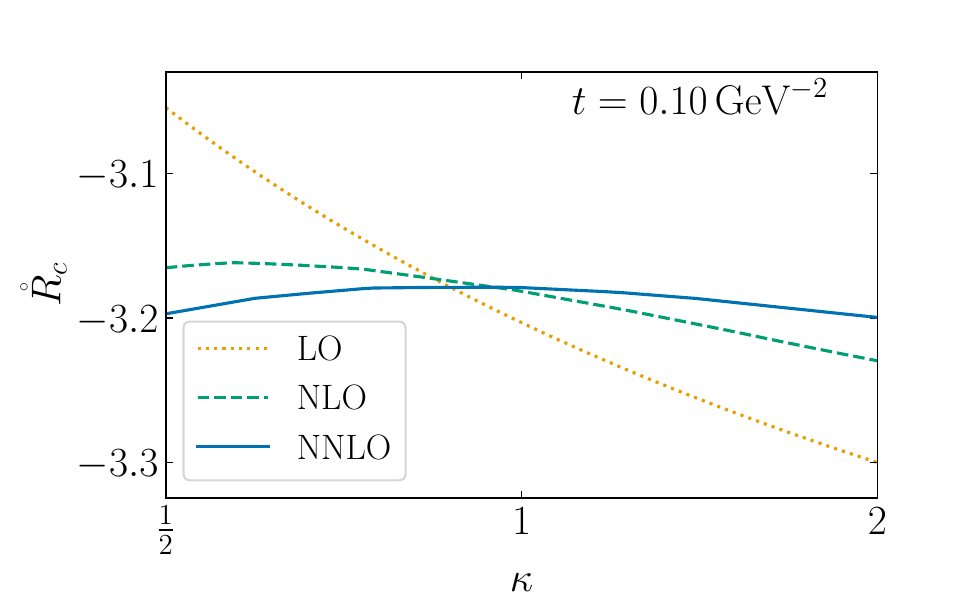} &
      \includegraphics[width=0.47\textwidth]{%
        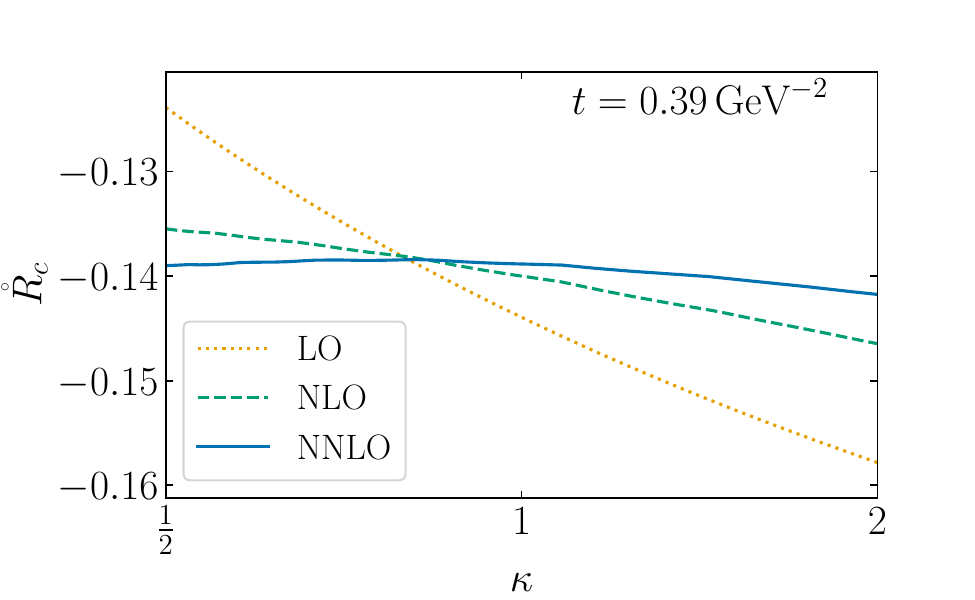}\\
      (a)& (b) \\ 
      \includegraphics[width=0.47\textwidth]{%
        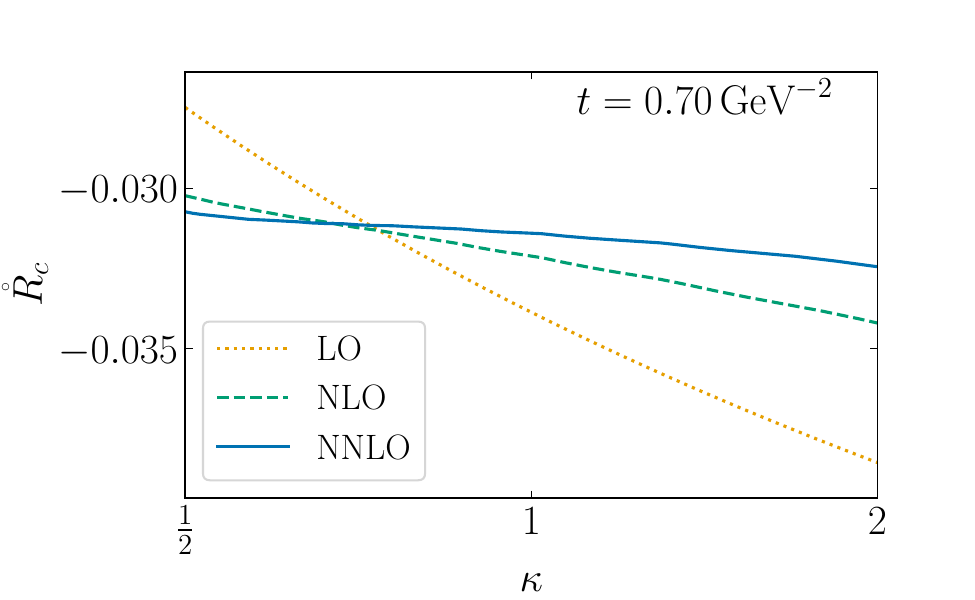}
      &
      \includegraphics[width=0.47\textwidth]{%
        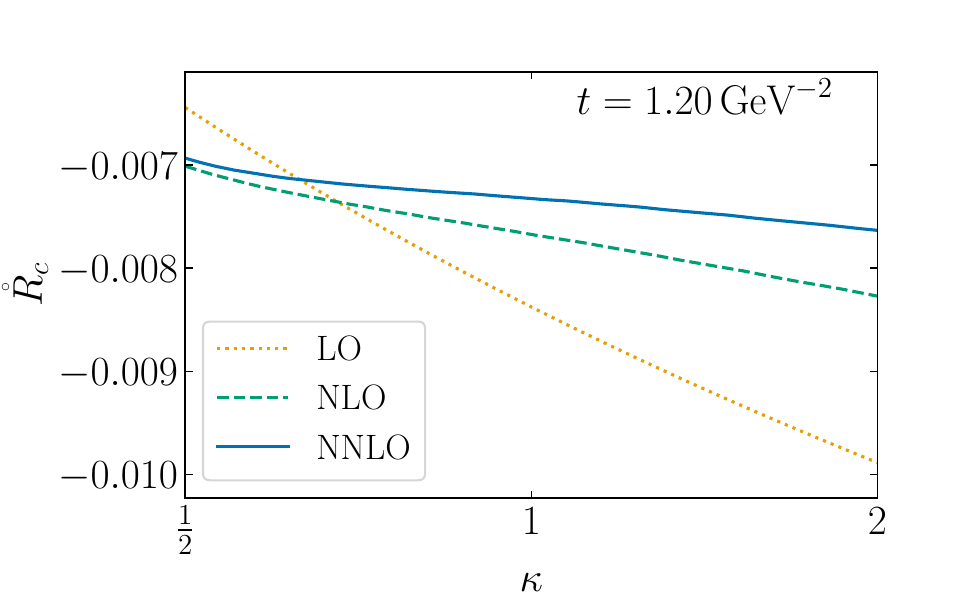} \\
      (c) & (d)
    \end{tabular}
    \caption{$\mathring{R}_c$ plotted across a range of
    $\kappa=\mu/\mu_{\text{int}}$ for various values of the flow time $t$.
    \label{fig:rbMuComp}}
  \end{center}
\end{figure}


While both for $\mathring{S}_f$ and $\mathring{R}_f$, a clear reduction of the
scale uncertainty towards higher orders is observed, we should mention a
qualitative difference between these two cases which is apparent from
\cref{fig:SMSCOMPHalf}. For $\mathring{S}_f$, one observes a throat in the
\lo\ data at around $t\approx 0.4\,\text{GeV}^{-2}$, both for $f=c$ and
$f=b$. It occurs because the $\mathcal{O}(a_s)$ coefficient to the scale
variation vanishes due to a cancellation between the $s_0$ and $s_0^{\prime}$
contributions, cf.~\cref{eq:Resu:enos}. We consider this in more detail in
\cref{fig:raMuComp} for $f=c$. Defining $\kappa = \mu/\mu_\text{int}$, it
shows $\mathring{S}_c(t)$ for four different values of the flow time,
$t\in\{0.1,~0.39,~0.7,~1.2\}\,\text{GeV}^{-2}$.  In each case, the reduction
in scale dependence of the observable is most apparent between \lo{} and
\nlo{}, with a slight but noticeable improvement at \nnlo{}.  Note that across
the throat, the direction of the scale variation changes; at the throat, the
\lo\ estimate of the uncertainty band is disjoint from those at higher orders.

For comparison, we provide the analogous plots for $\mathring{R}_f$ in
\cref{fig:rbMuComp}. In this case, the qualitative scale dependence hardly
changes between the different values of $t$, in keeping with the monotonous
scale variation bands observed in \cref{fig:SMSCOMPHalf}\,(c) and~(d).


\begin{figure}[!ht]
  \begin{center}
    \begin{tabular}{cc}
      \includegraphics[width=0.47\textwidth]{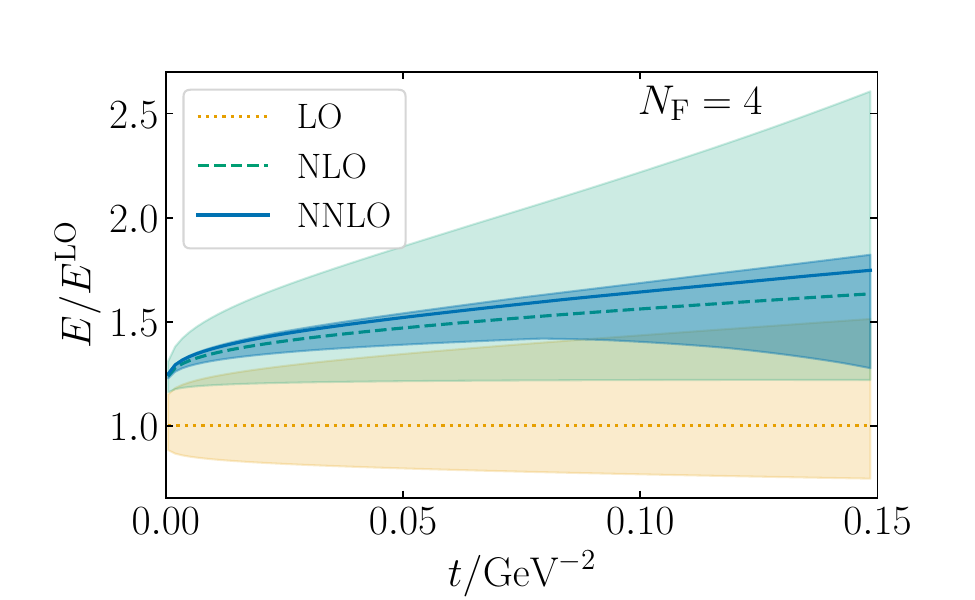} &
    \includegraphics[width=0.47\textwidth]{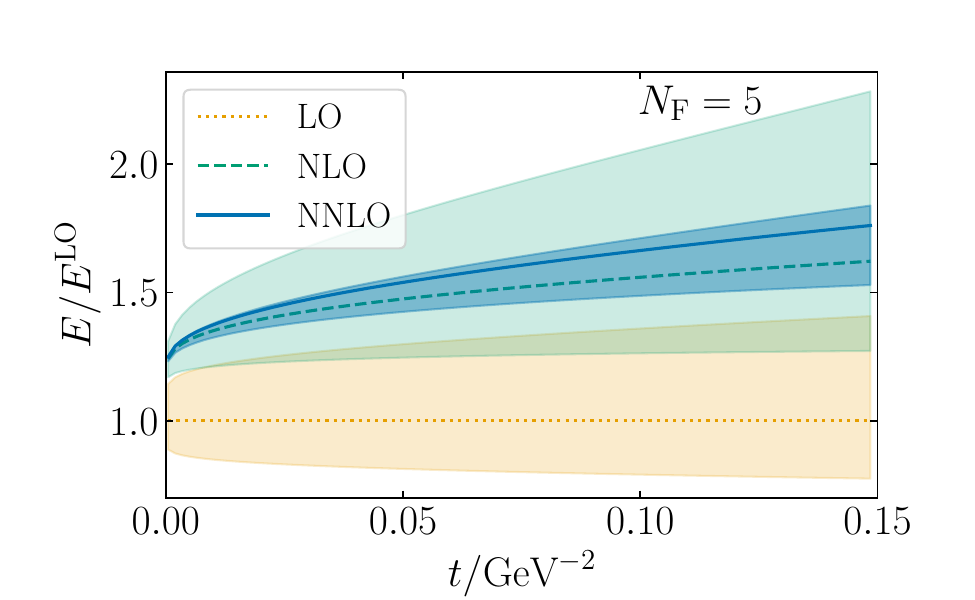}\\
    (a) & (b)\\
    \end{tabular}
    \caption{$E/E^\text{\lo}$ for (a)~$\nf=4$ (massive charm) and
      (b)~$\nf=5$ (massive bottom) at \nlo{} and \nnlo{}, with an envelope
      error formed by scale variation.\label{fig:EMSCOMPHalf}}
  \end{center}
\end{figure}


\begin{figure}[!ht]
  \centering
  \begin{center}
    \begin{tabular}{cc}
      \includegraphics[width=0.5\textwidth]{%
        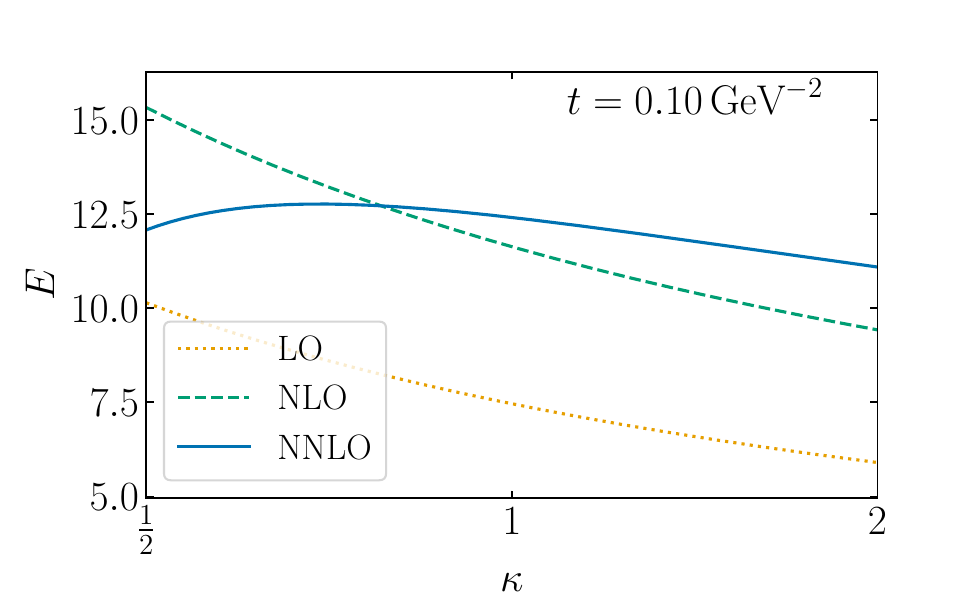} &
      \includegraphics[width=0.5\textwidth]{%
        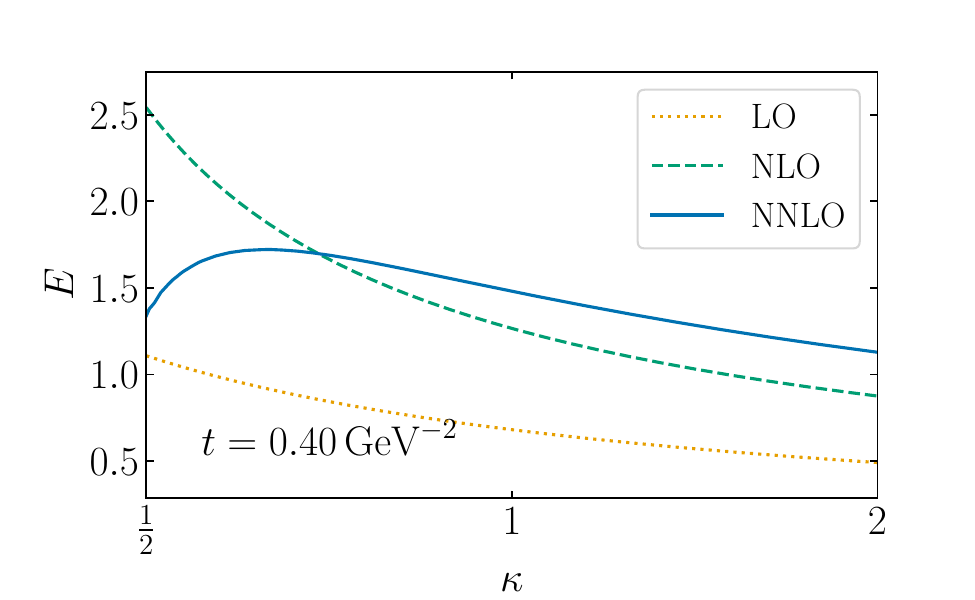} \\
      (a)& (b)
    \end{tabular}
  \caption{$E$ for $\nf=5$ plotted across a range of
    $\kappa=\mu/\mu_{t}$ for
    (a)~$t=0.1\,\text{GeV}^{-2}$ and
    (b)~$t=0.4\,\text{GeV}^{-2}$. \label{fig:EtMuComp}}
  \end{center}
\end{figure}


We can perform a similar analysis on $E(t)$. However, in this case we decided
to use $\mu_t$ rather than $\mu_\text{int}$ as the central scale, because the
quark masses play only a secondary role of this quantity. This is apparent
from the fact that, on the one hand, the \lo\ expression for $E(t)$ is
independent of the quark mass, and the mass effects at higher orders are
significantly smaller than for the quark operators $\mathring{S}_f(t)$ and
$\mathring{R}_f(t)$.  Due to this choice of the central scale, the range of
perturbative reliability in $t$ is diminished though.  In
\cref{fig:EMSCOMPHalf} we restrict the flow time to $t\leq
0.15\,\text{GeV}^{-2}$, corresponding to $\mu_t\geq 1.37$\,GeV, where the
scale uncertainty at \nnlo{} starts to increase, signaling the end of
perturbative reliability.  This can also be seen in \cref{fig:EtMuComp} which
considers the scale variation for a which uses larger $t$ values.


\subsection{Effects from a second quark mass}\label{sec:second}

As discussed above, up to now we have only considered a single quark as
massive, while the mass of the $\nf-1$ remaining ones was set to zero.  In the
remainder of this section, let us study the validity of this approximation by
allowing for a second massive quark.  For this discussion, we use $\mu_t$ as
the central scale, cf.\ \cref{eq:res:muint}.

Additional massive quarks occur as insertions into gluon propagators, and
higher-order corrections to that.  For $R_f(t)$ and $S_f(t)$ such diagrams
only appear at \nnlo{}, while for $E(t)$ they occur at \nlo{} in diagrams such
as that shown in \cref{fig:SpecFeynmanDiagrams}~(d). Our main concern here is
to keep both $m_c$ and $m_b$ different from zero. Since in $E(t)$ the masses
occur symmetrically, to a good approximation one can obtain the mass effects
for two massive quarks by simply adding the mass effects of
\cref{fig:EMSMLCOMP}\,(a) and~(b). This neglects effects arising from the
different values of $\nf$ in both plots, as well as \nnlo\ contributions from
diagrams which contain loop insertions from both massive quarks.


\begin{figure}[!ht]
  \begin{center}
    \begin{tabular}{cc}
      \includegraphics[width=0.47\textwidth]{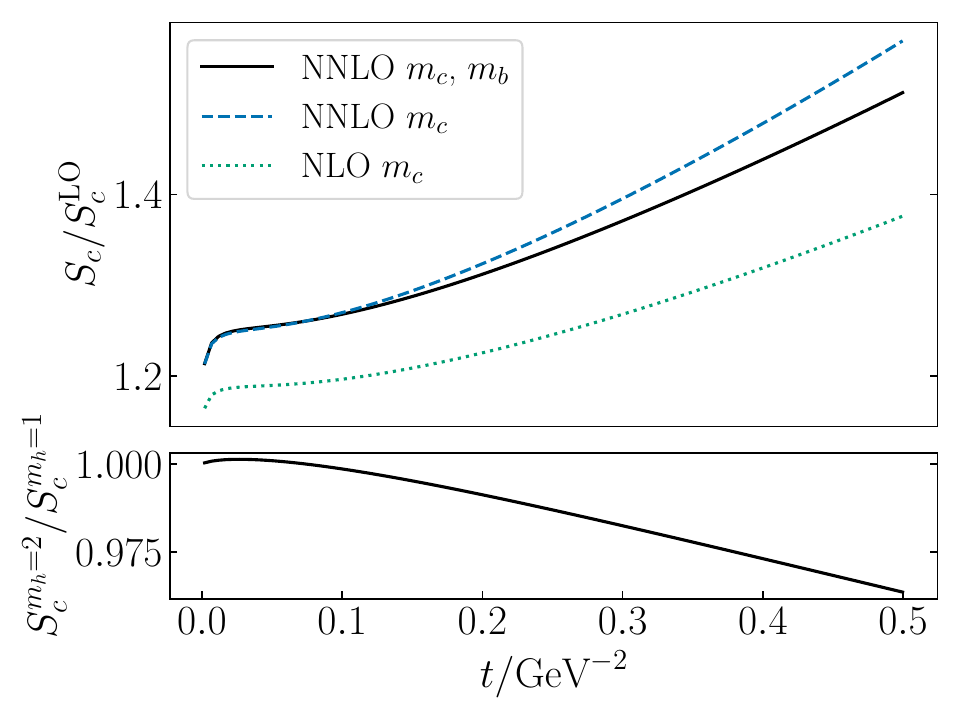} &
      \includegraphics[width=0.47\textwidth]{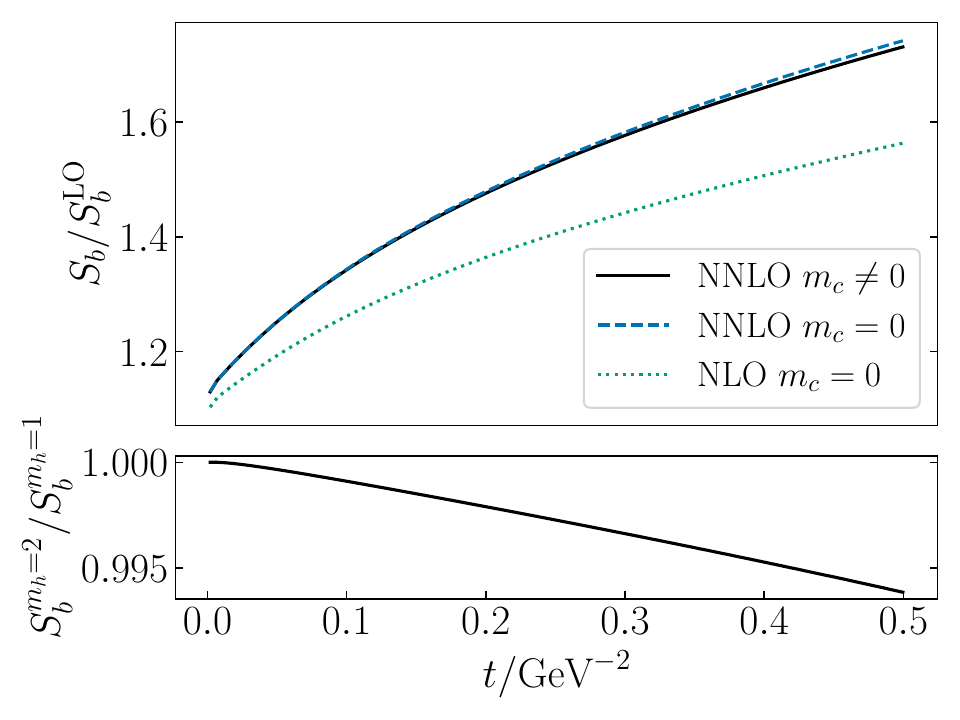} \\
      (a) & (b)
    \end{tabular}
    \caption{Effects of (a)~a massive bottom quark on $\mathring{S}_c$, and
      (b)~a massive charm quark on $\mathring{S}_b$. We have set 
      $\nf=5$ in both cases.}\label{fig:XIXIM2}
  \end{center}
\end{figure}


Concerning $\mathring{S}_f$ and $\mathring{R}_f$, the two massive quarks do
not appear symmetrically and one needs to distinguish $f=c$ and $f=b$.  We
study these two cases for $\mathring{S}_f$ explicitly in \cref{fig:XIXIM2}.
It compares the \two-mass case to the \lo{}, \nlo{} and \nnlo{} results for
one massive quark. The change due to the additional quark masses clearly
represents a sub-dominant contribution when compared with the jump from \nlo{}
to \nnlo{}. Within the displayed range of the flow time, the contribution from
the additional quark mass is less than the scale uncertainty in both cases,
cf.\,\cref{fig:SMSCOMPHalf}. However, while the charm-quark mass effects on
$\mathring{S}_b$ are at the sub-percent level, they do reach around 3\% for
the presence of a massive bottom quark in $\mathring{S}_c$.  We find similar
results for $\mathring{R}_f$, but refrain from showing them here.




\section{Conclusions}\label{sec:conclusions}

Using the \gff\ within perturbation theory, we have presented results for the
vacuum expectation values for phenomenologically relevant quark and gluon
operators in \qcd.  These have been computed numerically for one massive and
$\nl$ massless quarks, with the flow-time integrals being evaluated for a set
of points in the range $0.001<m^2t\leq63$. We provide the numerical data in an
ancillary file accompanying this paper.

As discussed in \citere{Takaura:2025pao}, two of these quantities, $S_f(t)$
and $R_f(t)$, provide the potential to extract renormalized quark masses by
comparing the perturbative results to lattice measurements. In the small-mass
approximation, radiative corrections to these quantities have been obtained
through \nnlo. Since they were found to be substantial, it is important to
control these corrections also in the case of massive quarks. Compared to the
leading-order mass effects, however, we find that the radiative corrections on
$S_f$ and $R_f$ are very modest. The perturbative prediction for these
quantities is therefore under very good control.

The quantity $E(t)$ can be used to define a renormalization scheme for the
strong coupling which can be implemented both on the lattice and within
perturbation theory~\cite{Luscher:2013vga}, for example, or to eliminate
perturbative renormalon contributions in predictions for physical
observables~\cite{Beneke:2025hlg}.  Computations on the lattice require finite
quark masses, therefore a consistent comparison to perturbation theory
requires the latter to include quark masses as well. In \cref{fig:EMSMLCOMP},
we showed the change due to massive quarks can be relevant, particularly at
\nnlo{}.

This project has been focused on the computation of the mass effects of
several \vevs{} which are fundamental to \qcd\ in the gradient flow. Using the
numerical data supplied with this paper in an ancillary file, it is
straightforward to evaluate ratios as suggested in \citere{Takaura:2025pao}
for the extraction of the quark masses.  One may also use these data to
evaluate derivatives of these quantities \wrt\ the quark mass, as this
eliminates non-perturbative effects that could spoil the mass
determination. However, such a numerical derivative is quite sensitive to the
accuracy and density of the original data.  Alternatively, one could take the
derivatives before loop integration.  However, at \nnlo\ this will lead to
additional and significantly more complicated integrals. We therefore consider
this beyond the scope of this work.


\paragraph{Acknowledgments.}
We would like to thank Hiromasa Takura for comparing numerical results at
\nlo\ and helpful comments, as well as Jonas Kohnen, Fabian Lange, David
Mason, and K\'alm\'an Szab\'o for useful discussions.  This research was
supported by the Deutsche Forschungsgemeinschaft (\abbrev{DFG}, German
Research Foundation) under grant 396021762 -- \abbrev{TRR} 257.



\begin{appendix}

  \section{Ancillary file}\label{sec:anc}

  We provide numerical results for the components defined in
  \cref{sec:sres,sec:rres,sec:eres} as \verb|numpy| arrays in the ancillary
  \verb|python| file \verb|DataFile.py|. The naming scheme for the arrays
  encoding $\mathring{S}_f(t)$ is
  \begin{equation}\label{eq:anc:culm}
    \begin{aligned}
      s_{0}\ &\hat{=}\ \mbox{\texttt{s0RSOS}}\\
      s_{i,j}\ &\hat{=}\ \mbox{\texttt{s<i><j>RSOS}}\\
      s_{i,j,k}\ &\hat{=}\ \mbox{\texttt{s<i><j><k>RSOS}}\\
      s'_{0}\ &\hat{=}\ \mbox{\texttt{s0pRSOS}}\\
      s'_{i,j}\ &\hat{=}\ \mbox{\texttt{s<i><j>pRSOS}}\\
      s''_{0}\ &\hat{=}\ \mbox{\texttt{s0ppRSOS}}\\
    \end{aligned}
  \end{equation}
  and similar for $\mathring{R}_f(t)$ and $E(t)$. For $\mathring{S}_f$ and
  $\mathring{R}_f$, these are \one-dimensional arrays of length 249; the
  corresponding values of $m^2t$ are given in the \verb|numpy| array
  \verb|m2tSR|. The arrays for $E(t)$ have length 200; the corresponding
  values for $m^2t$ are provided in the \verb|numpy| array \verb|m2tE|.

  For example, plotting $s'_{10}$ vs.\ $m^2t$ corresponds to the pseudo code
  \verb|plot(m2tSR,s10pRSOS)|. The analogous command for plotting
  $e_{2,0,0}$ is \verb|plot(m2tE,e200RSOS)|.
  
\end{appendix}


\bibliographystyle{utphys}
\bibliography{literatur}


\end{document}